\documentclass[nofootinbib, 10pt, twocolumn, superscriptaddress, notitlepage, tightenlines, aps,prb]{revtex4-1} 

\usepackage{tabstackengine}
\stackMath
\makeatletter
\renewcommand\TAB@delim[1]{\scriptstyle#1}
\makeatother
\setstackgap{S}{2pt}

\usepackage[sort&compress]{natbib}
\usepackage[utf8]{inputenc}
\usepackage[english]{babel}
\usepackage[mathscr]{euscript} 

\DeclareFontFamily{OT1}{pzc}{}
\DeclareFontShape{OT1}{pzc}{m}{it}{<-> s * [1.10] pzcmi7t}{}
\DeclareMathAlphabet{\mathpzc}{OT1}{pzc}{m}{it}

\usepackage{lipsum}
\usepackage[sc]{mathpazo}
\usepackage[T1]{fontenc}
\usepackage{amsmath}
\usepackage{relsize}
\usepackage{bigints}
\usepackage[usenames, dvipsnames]{xcolor}
\usepackage{hyperref}
\hypersetup{urlcolor=RoyalBlue, colorlinks=true,citecolor=RoyalBlue,linkcolor=OrangeRed} 
\usepackage{graphicx}
\usepackage{dcolumn}
\usepackage{bm}
\usepackage{amssymb}
\usepackage{comment}
\usepackage{amsthm}
\usepackage{booktabs}
\usepackage{epstopdf}
\usepackage{float}
\usepackage{tikz, pgfplots}
\usetikzlibrary{matrix}
\usepgfplotslibrary{fillbetween}
\usetikzlibrary{calc, automata, chains, arrows.meta, quotes}
\usepackage{enumitem}

\usepackage{thmtools}
\usepackage{thm-restate}

\usepackage[capitalise]{cleveref} 

\newcommand{\be}{\begin{eqnarray}}
\newcommand{\ee}{\end{eqnarray} }
\newcommand{\benn}{\begin{eqnarray*}}
\newcommand{\eenn}{\end{eqnarray*}}
\newcommand{\txt}{\textrm}

\newcommand{\D}{\txt{d}} 
\newcommand{\bpr}{\begin{proof}}
\newcommand{\epr}{\end{proof}}   
\newcommand\myeq{\mathrel{\overset{\makebox[0pt]{\mbox{\normalfont\tiny $L\rightarrow\infty$}}}{=}}}

\newcommand\mysimeqDeltat{\mathrel{\overset{\makebox[0pt]{\mbox{\normalfont\tiny $\Delta t \ll 1$}}}{\simeq}}}
\newtheorem{theorem}{Theorem}

\newtheorem*{theorem*}{Theorem}
\newtheorem*{corollary*}{Corollary}

\newtheorem{assumption}{Assumption}
\newtheorem{definition}{Definition}
\newtheorem{definition*}{Definition}[section]
\newtheorem{eg}{Example}

\newtheorem{lemma}{Lemma}
\newtheorem{corollary}{Corollary}

\newtheorem*{prop*}{Proposition}

\DeclareMathOperator*{\argmax}{arg\,max}

\begin{document}
\title{Work fluctuations due to partial thermalizations in two-level systems}
\author{Maria Quadeer}
\affiliation{\small{Centre for Quantum Software and Information, School of Computer Science, University of 	Technology Sydney, NSW 2007, Australia}}
\author{Kamil Korzekwa}
\affiliation{\small{Faculty of Physics, Astronomy and Applied Computer Science, Jagiellonian University, 30-348 Kraków, Poland}}
\author{Marco Tomamichel}
\affiliation{\small{Department of Electrical and Computer Engineering \& Centre for Quantum Technologies, National University of Singapore, Singapore 119077.}}

\begin{abstract}                 
   We study work extraction processes mediated by finite-time interactions with an ambient bath---\emph{partial thermalizations}---as continuous time Markov processes for two-level systems. Such a stochastic process results in fluctuations in the amount of work that can be extracted and is characterized by the rate at which the system parameters are driven in addition to the rate of thermalization with the bath. We analyze the distribution of work for the case where the energy gap of a two-level system is driven at a constant rate. We derive analytic expressions for average work and lower bound for the variance of work showing that such processes cannot be fluctuation-free in general. We also observe that an upper bound for the Monte Carlo estimate of the variance of work can be obtained using Jarzynski's fluctuation-dissipation relation for systems initially in equilibrium. Finally, we analyse work extraction cycles by modifying the Carnot cycle, incorporating processes involving partial thermalizations and obtain efficiency at maximum power for such finite-time work extraction cycles under different sets of constraints.
\end{abstract}

\date{\today}
\maketitle
\section{Introduction}\label{sec:intro}
A standard thermodynamic setting comprises of large systems with relatively short relaxation times wherein fluctuations in values of extensive quantities such as \emph{work}, that follow the law of large numbers, are negligible and one only cares about averages~\cite{Herbert_thermo, Jarzynski_review}. Work is essentially a deterministic quantity in such scenarios. Non-equilibrium thermodynamics, on the other hand, is the study of \emph{fluctuations} in work as one departs from the standard thermodynamic setting. Small systems with long relaxation times make the study of fluctuations inevitable since these are no longer just statistical noise. Within the framework of non-equilibrium statistical mechanics, fluctuations have been characterised using \emph{fluctuation theorems}~\cite{Jarzynski_PRL1997, Crooks} that play a key role in the control and study of biomolecular processes such as the folding of proteins~\cite{Dobson2003}. Single molecule experiments involving stretching of biomolecules under external forces are a ripe avenue for the study of non-equilibrium phenomena~\cite{Alemany&Retort2010, Bustamante_2005, fluctuations-biomolecular-review, Liphardt1832}. Another complementary approach to non-equilibrium thermodynamics is the incipient field of one-shot statistical mechanics~\cite{Garner_2018, Aberg-work-extraction_2013, Dahlsten_2011, lidia_2011thermodynamic,Faist_2015} which draws techniques from one-shot information theory~\cite{tomamichel2015quantum, renyi1961, renner-wolf2004, renner_phd-thesis2005} to characterize processes that are far from equilibrium. And, it is in this non-equilibrium framework that work is analysed as a random variable.

The one-shot regime considers single instances of the task at hand instead of looking at an ensemble average. Fluctuations in work have been studied in this regime in Ref. \onlinecite{Aberg-work-extraction_2013} for a discrete classical model. The main question in consideration was what constituted truly work-like work extraction. The basic idea was that in order to define work for small systems in contrast to \emph{heat}, one should be able to extract a fixed amount of work from a fixed system configuration. The author showed that for a system that is initially not in equilibrium with the ambient bath even the optimal process (that achieves maximum average work output) results in fluctuations as large as the average work itself. Such an optimal process is comprised of a) energy level transformations (quenches) rendering the system effectively thermal, and b) \emph{reversible isothermal} processes~\cite{Aberg-work-extraction_2013}. The latter is manifestly fluctuation-free because the system equilibrates at each infinitesimal step of a reversible process and equilibration washes away the fluctuations. Of course this is only possible since the time scale over which one performs the energy level transformations (during the isothermal process) is much larger than the relaxation time of the system. The question is what happens to these fluctuations when the system only \emph{partially} thermalizes, i.e. when the system is driven externally over shorter time periods in comparison to its relaxation time scale.

The relaxation towards equilibrium can be studied within the framework of \emph{collision models} which have been used to study open quantum system dynamics~\cite{CollisionModel_OpenQuantumDynamics2005, CollisionModel_ThermalizingQuantumMachines_PRL2002, Collisionmodels_quantumoptics}. Within these models the bath is treated as a composition of smaller non-interacting particles that are copies of the system in the thermal Gibbs state. Such a process of thermalization has been studied in Ref.~\onlinecite{CollisionModel_ThermalizingQuantumMachines_PRL2002} for a qubit in contact with a bath composed of non-interacting qubits. For a qubit in a general quantum state interacting with an ambient bath, thermalization was shown to be a two-component process comprising of decoherence and dissipation. And, a functional dependence on time was obtained for both of these processes. In the present work, we are interested in fluctuations in processes involving partial thermalizations for classical two-level systems and will limit ourselves to states diagonal in the energy eigenbasis. And, on this note, we come to the question we posed above regarding fluctuations in work for processes involving partial thermalizations.

A numerical study on this question was undertaken in Ref.~\onlinecite{Dhar_PRE2005} where the authors studied a single Ising spin driven by an external magnetic field. They obtained work distributions using Monte Carlo simulations of the processes for different driving rates. The authors found that such processes have broad work distributions with significant probability for processes with negative dissipated work in general. They also verified work fluctuation theorems~\cite{Crooks, Jarzynski_PRL1997} and derived analytic expressions for the distribution of work when the spin's energy gap was driven by the external field in the slow and fast limits. Another recent work~\cite{Baumer2019imperfect} looked into the same problem but ignored the dependence of the system-bath interaction on time. We will discuss this further in \cref{sec:Model}. 

In this paper, we investigate similar work extraction processes involving partial thermalizations for a single classical two-level system driven by an external magnetic field changing linearly in time. We derive an analytic expression for the average work yield of such a process as a function of the total time, $\tau$. This expression reduces to the average work outputs of the corresponding adiabatic and isothermal processes in the $\tau\rightarrow0$ and $\tau\rightarrow\infty$ limits respectively. Next, in an attempt to characterize fluctuations in the average work yield, we provide a lower bound for the variance of work as a function of the total time duration of the process. This lower bound is saturated in the adiabatic and the isothermal limits thereby reproducing the result that isothermal processes  are deterministic as was shown in ~Ref.~\onlinecite{Aberg-work-extraction_2013}. Even though an analytical expression for the variance of work seems intractable, we employ Jarzynski's fluctuation-dissipation relation \cite{Jarzynski_PRL1997} to compare the dissipation in work $(\tau < \infty)$ with our estimate of variance obtained by performing Monte Carlo simulations. We find that for a two-level system initially in equilibrium with the bath, the fluctuation-dissipation relation provides a good approximation which becomes exact as $\tau$ becomes large. This was also noted in Ref.~\onlinecite{Dhar_PRE2005}. Finally, we investigate finite-time work extraction cycles inspired by the Carnot cycle replacing the ideal isothermal reversible processes with the realistic ones involving partial thermalizations. We then numerically optimize the power output of such finite-time work extraction cycles over different sets of constraints and parameters keeping the time period of the cycles fixed and provide comparisons between those scenarios.

This paper is divided into six sections. In \cref{sec:Model} we define finite-time work extraction processes involving partial thermalizations as a Markov process and describe the microscopic model for a two-level system. In \cref{sec:results} we derive the analytical results and in \cref{sec:numerical_results} we discuss the results from Monte-Carlo simulations. We then analyze finite-time heat engines in \cref{sec:application} and summarize our work in \cref{sec:Summary}.
\section{Model}\label{sec:Model}
Given an ambient bath at temperature $T_h$ and a two-level system such that its energy gap $\delta$ can only be driven within a fixed range between $\delta_{min}$ and $\delta_{max}$ (for example by an external magnetic field), let us assume that the time period of the external driving is much shorter compared to the relaxation time of the system interacting with the bath. Furthermore, we assume that the spectral density of the bath is constant over the given range of values of the energy gap. And, without loss of generality, we assume that the ground state energy is zero. Note that for a two-level system,  given an energy gap $\delta$, one can always define a temperature $T$ such that the occupation probability of the excited state of the system is given by the corresponding Gibbs distribution at temperature $T$. We choose time as the independent quantity under these settings and denote it by the continuous variable $t$. With these initial constraints, the question is how can one extract work. To this end, we study finite-time work extraction processes involving  \emph{partial thermalizations}. Partial thermalization encapsulates a finite time restriction for the system's equilibration with the bath and can be studied by considering a randomized model of interaction between the two---a collision model~\cite{CollisionModel_OpenQuantumDynamics2005, CollisionModel_ThermalizingQuantumMachines_PRL2002, Collisionmodels_quantumoptics}. Such models are based on the assumption that the bath is composed of smaller non-interacting particles that are copies of the system in the thermal Gibbs state. The system-bath interaction is then modelled as a sequence of collisions between the system and bath particles where each collision itself is considered to be a joint unitary on the system and the bath particle in question. The additional assumptions of the bath being initially uncorrelated and the system colliding with exactly one bath particle at a time result in a Markovian dynamics for the system which in turn can be translated to a Lindblad master equation in the continuous time limit \cite{CollisionModel_LME_ToddBrun, OpenSystem_CollisionModel_Ziman}. The process of partial thermalization was studied in Ref.~\onlinecite{CollisionModel_ThermalizingQuantumMachines_PRL2002} within the framework of a collision model and was shown to be composed of dissipation and decoherence for a general quantum state. For states that are diagonal in the energy eigenbasis thermalization simply amounts to dissipation and the state of a two-level system can be described by the occupation probabilities $p(t)$ for the excited state and $1-p(t)$ for the ground state. Denoting the thermal Gibbs occupation probability for the excited state by $\gamma(t)$, the thermalization process is given by the following equation as per Ref.~\onlinecite{CollisionModel_ThermalizingQuantumMachines_PRL2002}:
\be \label{eq:pth_dissipation}
p(t) = e^{-\kappa t} p(0) + \big(1-e^{-\kappa t}\big)\gamma(t),
\ee 
where $\kappa$ is the thermalization rate (dissipation), the inverse of the relaxation time $T_1$~\cite{CollisionModel_ThermalizingQuantumMachines_PRL2002}. One can interpret $1-e^{-\kappa t}$ as the probability of collision between the qubit and a bath particle, denoting it by $\lambda$. The case $\lambda =1$ corresponds to exact thermalization and $\lambda=0$ corresponds to no thermalization. Thus, for sufficiently short interaction times $\Delta t$, the probability $\lambda$ with which the system interacts with the bath particles (and thermalizes) is linear in $\Delta t$, i.e. $\lambda = \kappa \Delta t$.
Microscopically, partial thermalization is a time-dependent Markov process on a finite state space---the ground and excited states of our two-level system. The system (with energy gap $\delta(t)$ at time t) interacts with the bath for a time $\Delta t$ and with probability $\kappa \Delta t$ it collides with a bath particle. If the system thermalizes then it can change its state such that the occupation probability for the excited state is $\gamma_h\big(\delta(t+\Delta t)\big)$, the thermal Gibbs weight associated with the excited state $\delta(t+\Delta t)$ for the bath temperature $T_h$. Work is done when the system is in the excited state and its energy gap changes from $\delta(t)$ to $\delta(t+\Delta t)$.

We can thus build a work extraction process as per ~Ref.~\onlinecite{Aberg-work-extraction_2013} where one performs a series of infintesimal level transformations followed by \emph{partial} thermalizations instead of thermalizing the system exactly. A discrete version of such a process at a given time $t$ is therefore composed of a series of two steps:
	\begin{itemize}
    \item[]1. Level transformation: changing the energy gap $\delta (t)$ by an infinitesimal amount to $\delta(t+\Delta t)$ keeping the occupation probabilities fixed.
    \item[]2. Partial thermalization: changing the state of the system such that with probability $1-\kappa \Delta t$ it stays in the same state, while with probability $\kappa \Delta t$ it thermalises with respect to the bath at temperature $T_h$.  
    \end{itemize}
This process is depicted as a Markov diagram in \cref{fig:Markovchain} below. 
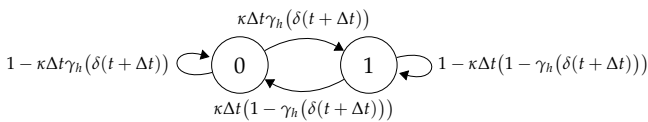
\begin{figure}[H] 
\vspace{0.5cm}
\hspace{3cm}
	\begin{tikzpicture}[start chain = going right,
     -Triangle, every loop/.append style = {-Triangle}, transform canvas={scale=0.95}]
     \foreach \i in {0,1} 
       \node[state, on chain]  (\i) {\i};
    \foreach \i in {0} 
    {
       \draw let \n1 = { int(\i+1) } in
       (\i)  edge[bend left, "\scriptsize $\kappa \Delta t\gamma_h\big(\delta(t+\Delta t)\big)$" ] (\n1)
        (\n1) edge[bend left, "\scriptsize $\kappa \Delta t\big(1-\gamma_h\big(\delta(t+\Delta t)\big)\big)$"] (\i);
     }
     \draw    (0)  edge[loop left, "\scriptsize $1-\kappa \Delta t\gamma_h\big(\delta(t+\Delta t)\big)$"]   (0);
     \draw    (1)  edge[loop right, "\scriptsize $1-\kappa \Delta t\big(1-\gamma_h\big(\delta(t+\Delta t)\big)\big)$"]  (1);s     
    
	\end{tikzpicture} 
	\vspace{0.75cm}
	\caption{Markov chain representing discrete partial thermalization at time $t+\Delta t$. The states 0 and 1 denote the corresponding ground and excited states of the two level system.}
	\label{fig:Markovchain}
\end{figure}
The time evolution of a continuous partial thermalization process can be seen as a limiting case of the discrete Markov process above as shown in the following lemma.
\begin{figure*}[t]
\begin{center}
    \begin{tikzpicture}
     \matrix (m)[
       matrix of math nodes,
       nodes={circle,draw},
       minimum size=0.75cm,
       inner sep=1mm,
       column sep=25mm,
       row sep=10mm
     ]
       {
         1 & 1 & 1 \\
         0 & 0 & 0 \\
       };
       \path[->, font=\scriptsize]
       (m-2-1) edge[RubineRed] node[rotate=0, xshift=0cm, yshift=0.2cm]{$1-\kappa \Delta t\gamma_h(\Delta t)$} (m-2-2)
       
       (m-1-2) edge[dashed, ForestGreen] (m-1-3)
       
       (m-1-2) edge[dashed, ForestGreen] (m-2-3)
       
       (m-2-1) edge[ForestGreen] node[rotate=30, xshift=0cm, yshift=0.2cm]{$\kappa \Delta t\gamma_h(\Delta t)$} (m-1-2)
       
       (m-2-2) edge[dashed, RubineRed] (m-1-3)
       
        (m-2-2) edge[dashed, RubineRed] (m-2-3)
       
       (m-2-1) node[yshift=-1cm]{\normalsize $t=0$}
       
       (m-2-2) node[yshift=-1cm]{\normalsize $t=\Delta t$}
        
       (m-2-3) node[yshift=-1cm]{\normalsize $t=\tau$};
       
%
       
   \end{tikzpicture}
   \label{fig:eg1-Markovchain}
\end{center} 
\end{figure*}
\begin{figure*}[t] 
  \centering
    \begin{tabular}{|c|c|c|c|c|c|} 
      ~~state at $t=0$~~ & ~~$W^{t=\Delta t}$~~ & ~~state at $t=\Delta t$~~ & ~~$W^{t=\tau}$~~ & ~~~$W$~~~ & ~~$\Pr{[W]}$~~\\
      \hline
      0 & 0 & 0 & 0 & \textcolor{RubineRed}{0} & $(1-p_0)\big(1-\kappa \Delta t\gamma_h(\Delta t)\big)$\\
      \hline
      0 & 0 & 1 & $\epsilon_2$ & \textcolor{ForestGreen}{$\epsilon_2$} & $(1-p_0)\kappa \Delta t\gamma_h(\Delta t)$\\
      \hline
      1 & $\epsilon_2$ & 0 & 0 & \textcolor{ForestGreen}{$\epsilon_2$} & $p_0\kappa \Delta t\big(1-\gamma_h(\Delta t)\big)$\\
      \hline
      1 & $\epsilon_2$ & 1 & $\epsilon_2$ & \textcolor{Cyan}{$2\epsilon_2$} & $p_0\big(1-\kappa \Delta t\big(1-\gamma_h(\Delta t)\big)\big)$
     \end{tabular}
  \caption{\textit{Occupation of the ground state is designated by $0$ and that of the excited state by $1$. Each row in the table corresponds to a specific path. The Markov chain above shows two specific paths. The first and third columns denote the state of the system at the beginning of each step while the second and the fourth columns denote the work done at the corresponding steps. $W$ is the sum of work done at each of the steps along a given path and $\Pr[W]$ is the probability for each value of $W$ which is obtained using \cref{fig:Markovchain}.}}
   \label{tab:table6}
\end{figure*}  
\begin{lemma}[Continuous time partial thermalization]\label{lemma:dp(t)/dt}
	Given a two-level system in the presence of a hot ambient bath at temperature $T_h$, the occupation probability $p$ for the excited state evolves according to the following equation during a general partial thermalization process characterized by short system-bath interaction times:
\be\label{eq:lemma:PTh_dp/dt}
\frac{\D p(t)}{\D t} = \kappa\Big(\gamma_{h}\big(\delta(t)\big) - p(t)\Big),
\ee 
where $\gamma_{h}\big(\delta(t)\big)=\frac{1}{1+e^{\delta(t)/T_h}}$, the Gibbs weight associated with the excited state $\delta(t)$.
\end{lemma}
\bpr
According to \cref{fig:Markovchain}, the total probability of being in the excited state $p(t+\Delta t)$ at time step $t+\Delta t$ can be obtained using the law of total probability:
\be
p(t+\Delta t)&=& p_{01}(t+\Delta t)\big(1- p (t)\big) + \notag \\
&~&p_{11}(t+\Delta t)p(t),
\ee
where $p_{01}(t+\Delta t)$ is the conditional probability for the system to be in the excited state at time $t+\Delta t$ when it was in the ground state at time $t$, and $p_{11}(t+\Delta t)$ is the conditional probability for the system to be in the excited state at time $t+\Delta t$ when it was in the excited state at time $t$. Plugging in the corresponding expressions using \cref{fig:Markovchain} we have 
\be
p(t+\Delta t)=(1-\kappa \Delta t)p(t) +\kappa \Delta t \gamma_{h}\big(\delta(t+\Delta t)\big).
\ee
Re-arranging the terms we obtain 
\be \label{eq:refute_Baumer}
p(t+\Delta t) - p(t)&=&\kappa\Delta t\Big(\gamma_{h}\big(\delta(t+\Delta t)-p(t)\big)\Big),
\ee
which after dividing by $\Delta t$ reduces to \eqref{eq:lemma:PTh_dp/dt} in the limit $\Delta t\rightarrow 0$.
\epr 	
As an aside, we would like to make a comment on the model of partial thermalization as in~Ref.~\onlinecite{Baumer2019imperfect}. The authors consider a situation where the probability of interaction between the system and the bath is fixed. If we were to do the same then we would have to replace $\kappa \Delta t$ by a constant, let us say $\lambda$. Then \eqref{eq:refute_Baumer} would yield
\be 
p(t+\Delta t) - p(t)&=&\lambda\Big(\gamma_{h}\big(\delta(t+\Delta t)-p(t)\big)\Big),
\ee 
which in the limit $\Delta t\rightarrow 0$ would simply give 
\be 
p(t) = \gamma_{h}\big(\delta(t+\Delta t)\big).
\ee 
This implies that the system would be in the thermal Gibbs state at each infinitesimal step of the process. Naturally, one would obtain the same result as an isothermal reversible with no fluctuations. 

Now that we have a general model of partial thermalization, we make the following assumption about the rate at which the energy gap $\delta(t)$ is driven in time. 
\begin{assumption}\label{assumption:1}
	The energy gap of a two-level system is driven at a constant rate, i.e.
	\be \label{eq:PTh_ddelta/dt}
		\frac{\D \delta}{\D t}=\textrm{constant}.
	\ee 
\end{assumption}	
This completes our model of a finite-time work extraction process with partial thermalizations. We are thus ready to answer the questions posed in \cref{sec:intro}, i.e. what is the average work yield of such processes and are they fluctuation-free. But, first we look at an example of a discretized version of this problem for an intuitive understanding of the underlying Markov process which would inform our derivations in \cref{subsec:lb-variance}.

\begin{eg}\label{eg:1}
	Given that $\epsilon = \delta_{max}-\delta_{min}$ is the available range over which we can vary $\delta$ as a function of time $t$, let us choose $\delta(0)=\delta_{max}$ and $p(0)$ to be some constant $p_0$. Let us denote the total time duration of the process by $\tau$. Then, $\delta(\tau)=\delta_{min}$ and $p(\tau)$ would be determined by the process itself. 
Using \cref{assumption:1} under the above boundary conditions, we find that $\D\delta/\D t = -\epsilon /\tau$. A work extraction process involving partial thermalization corresponds to a curve on the $\delta-p$ plane. One should thus think of the discretization of the process as a discretization of this curve. So, let us divide $\delta$ over the available range into $L=2$ equal steps. Then, the change in $\delta$ at each step is $\Delta \delta= -\epsilon /2\triangleq\epsilon_2$. This implies that $\Delta t = \tau /2$ for each step. As discussed earlier, each of these discrete steps is composed of a level transformation followed by a partial thermalization. Let us say that the system is in the ground state at $t=0$ as shown in \cref{tab:table6}. Thus, the work done in the first step during level tranformation would be zero. Next, the system is thermalised with respect to the hot bath with probability $\kappa \Delta t$. The work done during partial thermalization is dissipated as heat and thus its contribution is zero. Upon partial thermalization we might transition to the excited state or remain in the ground state. The two possible paths are shown in the Markov chain in \cref{tab:table6}. If we transition to the excited state then the work done would be $\epsilon_2$ during the level transformation in the second step. Finally, partial thermalization in the second step would again lead to two different paths (dashed lines in the Markov chain) each corresponding to zero work output. The probabilities corresponding to these paths add up to 1. The complete distribution of work can thus be obtained by going through all such paths as enlisted in the table in \cref{tab:table6}.
\end{eg}

\section{Analytical results}\label{sec:results}
In this section, we derive an expression for average work done during work extraction processes involving partial thermalizations (\cref{subsec:average-work}) and prove that they are \emph{not} fluctuation-free in general (\cref{subsec:lb-variance}).
\subsection{Average work}\label{subsec:average-work}


Let us denote the work done during a general thermodynamic process by the random variable $W$. As we discussed in \cref{eg:1}, work is done when a two-level system is in the excited state and a level transformation occurs. Note that depending upon whether we increase or decrease the energy gap, one would obtain negative or positive values of work $W$ corresponding to a net work gain or a net work cost. In this paper, we shall denote a net work gain by the random variable $W$ and refer to it as \emph{just} work, as a convention. Thus, the average work of a finite-time process where the energy gap changes from $\delta_{max}$ to $\delta_{min}$ along with partial thermalizations for a time $\tau$ would be
\be \label{eq:PTh_<W>_cts}
\mu_W(\tau) = - \bigintss_{\delta_{max}}^{\delta_{min}}\hspace{-5mm} p(\delta) \D \delta,
\ee 
where $p(\delta)$ is the probability of the system to be in the excited state when the energy gap is $\delta$. We will first list a few ingredients that would come in handy in deriving the main result, i.e. an expression for average work, \cref{theorem:average work}.
\begin{definition} Given the energy gap $\delta(t)$ of a two-level system at time $t<\tau$, we define the function
	\be \label{def:G}
		\mathcal{G}: t \mapsto - \mathlarger{\mathlarger{\mathlarger{\sum}}}_{n=1}^\infty \frac{\Big(-e^{-\frac{\delta(t)}{T_h}}\Big)^n}{\Big(\frac{n\epsilon}{\kappa\tau T_h}+1\Big)},
	\ee 
	where $\epsilon=\delta_{\max}-\delta_{\min}$, $\kappa$ is the thermalization rate, and $T_h$ is the temperature of the ambient bath.
\end{definition}
The function $\mathcal{G}$ is a monotone function in $t$. For $\delta$ monotonically decreasing in $t$, $\mathcal{G}$ monotonically increases. This follows by noting that $-e^{-\delta(t)}$ is also monotonically decreasing in $t$. We also make use of a few standard functions in the proofs that have been redefined in \cref{app:special_functions} for completeness.

\begin{lemma}[Time evolution of occupation probabilities under partial thermalization]\label{lemma:p(t)}
Given a two-level system that undergoes partial thermalization as per \cref{assumption:1} in the presence of a bath at temperature $T_h$ for a time $\tau$ such that its energy gap changes from $\delta_{max}$ to $\delta_{min}$, the probability of the system to be in the excited state at any time $0<t<\tau$ is 
\be \label{eq:lemma:p(t)}
p(t)&=&p_0 e^{-\kappa t} + \mathcal{G}(t)-e^{-\kappa t}\mathcal{G}(0),
\ee 	
where $\epsilon =\delta_{max}-\delta_{min}$, $p_0 = p(0)$ and $\delta(t) = \delta_{max} - \epsilon t /\tau$.
\end{lemma}
\bpr
Re-writing the differential equation for general partial thermalization processes, \eqref{eq:lemma:PTh_dp/dt}, we have
\be
\frac{\D p}{\D t} + \kappa p(t) = \kappa \gamma_h\big(\delta(t)\big),
\ee
which can be integrated along with the initial condition $p(0)=p_0$ to obtain 
\be\label{eq:lemma:p(t):1}
p(t) = p_0 e^{-\kappa t} + \kappa e^{-\kappa t}\int_0^t e^{\kappa t'} \gamma_h\big(\delta(t')\big) ~\D t'.
\ee
Given \cref{assumption:1} and the boundary conditions $\delta(0)=\delta_{max}$ and $\delta(\tau)=\delta_{min}$, we have 
\be\label{eq:lemma:p(t):3}
\delta(t)=\delta_{max} - \frac{\epsilon}{\tau} t,
\ee
where $\epsilon = \delta_{max}-\delta_{min}$. Plugging
$
\gamma_h\big(\delta(t)\big)=\frac{1}{1+e^{\delta(t)/T_h}}
$ and \eqref{eq:lemma:p(t):3} in \eqref{eq:lemma:p(t):1}, we obtain 
\be\label{eq:lemma:p(t):4}
p(t) = p_0 e^{-\kappa t} + \kappa e^{-\kappa t}\bigints_0^t \frac{e^{\kappa t'}}{1+e^{\frac{(\delta_{max} -\epsilon t'/\tau)}{T_h}}} ~\D t'.
\ee 
The integral above is given in terms of the Hypergeometric function as in \eqref{def:eq:general integral_hypergeometric}. Thus, 
\be\label{eq:lemma:p(t):5}
p(t)&=&p_0 e^{-\kappa t} + \kappa e^{-\kappa t}\Bigg\{\frac{e^{\big(\kappa + \frac{\epsilon}{\tau T_h}\big)t'-\frac{\delta_{max}}{T_h}}}{\kappa + \frac{\epsilon}{\tau T_h}}\times \notag \\
&~&_2F_1\Bigg(1, \frac{\kappa \tau T_h}{\epsilon}+1, \frac{\kappa \tau T_h}{\epsilon}+2; -e^{-\frac{(\delta_{max} -\epsilon t'/\tau)}{T_h}}\Bigg)\Bigg|_0^t \Bigg\} \notag \\
&=&p_0 e^{-\kappa t} + \kappa e^{-\kappa t}\Bigg\{\frac{e^{\big(\kappa + \frac{\epsilon}{\tau T_h}\big)t -\frac{\delta_{max}}{T_h}}}{\kappa + \frac{\epsilon}{\tau T_h}}\times \notag \\
&~&_2F_1\Bigg(1, \frac{\kappa \tau T_h}{\epsilon}+1, \frac{\kappa \tau T_h}{\epsilon}+2; -e^{-\frac{(\delta_{max} -\epsilon t/\tau)}{T_h}}\Bigg) -\notag \\
&~&\frac{e^{-\frac{\delta_{max}}{T_h}}}{\kappa + \frac{\epsilon}{\tau T_h}}~_2F_1\Bigg(1, \frac{\kappa \tau T_h}{\epsilon}+1, \frac{\kappa \tau T_h}{\epsilon}+2; -e^{-\frac{\delta_{max}}{T_h}}\Bigg)\Bigg\} \notag \\
&=&p_0 e^{-\kappa t} + \frac{\kappa\tau T_h}{\epsilon} \Bigg\{\frac{e^{-\frac{(\delta_{max} -\epsilon t/\tau)}{T_h}}}{\frac{\kappa \tau T_h}{\epsilon} +1}\times \notag \\
&~&_2F_1\Bigg(1, \frac{\kappa \tau T_h}{\epsilon}+1, \frac{\kappa \tau T_h}{\epsilon}+2; -e^{-\frac{(\delta_{max} -\epsilon t/\tau)}{T_h}}\Bigg) - \notag \\
&~&e^{-\kappa t}\frac{e^{-\frac{\delta_{max}}{T_h}}}{\frac{\kappa \tau T_h}{\epsilon} +1}\times \notag \\
&~&_2F_1\Bigg(1, \frac{\kappa \tau T_h}{\epsilon}+1, \frac{\kappa \tau T_h}{\epsilon}+2; -e^{-\frac{\delta_{max}}{T_h}}\Bigg)\Bigg\}.
\ee
Next, using \eqref{def:eq:hypergeometric_as_series} we can write 
\be\label{eq:lemma:p(t):6}
&~&\bigg(\frac{a z}{a +1}\bigg) ~_2F_1(1, 1+ a, 2+a; -z)\notag \\
&=&\bigg(\frac{a z}{a +1}\bigg)\mathlarger{\mathlarger{\sum}}_{n=0}^\infty \frac{n! (1+a)_n}{(2+a)_n} \frac{(-z)^n}{n!}\notag \\
&=& \bigg(\frac{a z}{a +1}\bigg)\mathlarger{\mathlarger{\sum}}_{n=0}^\infty \frac{(1+a)(2+a)\cdots (n+a)}{(2+a)(3+a)\cdots (n+a+1)} (-z)^n \notag\\
&=& -a  \mathlarger{\mathlarger{\sum}}_{n=0}^\infty \frac{(-z)^{n+1}}{(n+a+1)} \notag \\
&=& -a  \mathlarger{\mathlarger{\sum}}_{n'=1}^\infty \frac{(-z)^{n'}}{(n'+a)} \notag \\
&=& - \mathlarger{\mathlarger{\sum}}_{n'=1}^\infty \frac{(-z)^{n'}}{(\frac{n'}{a}+1)}.
\ee
Using \eqref{eq:lemma:p(t):6} we can write \eqref{eq:lemma:p(t):5} in terms of the function $\mathcal{G}$, \cref{def:G}, to obtain \eqref{eq:lemma:p(t)}.
\epr
Having listed the ingredients above, we are ready to derive the expression for average work.
\begin{theorem}[Average work]\label{theorem:average work} The average work done during a finite-time process by a two-level system such that its energy gap changes from $\delta_{max}$ to $\delta_{min}$ as per \cref{assumption:1} along with partial thermalizations with respect to a bath at temperature $T_h$ for a time $\tau$  is
 
\be
\mu_W(\tau)  &=& W_{iso}^{T_h} +  \frac{W_{ad}}{\kappa \tau}\big(1-e^{-\kappa \tau}\big) \notag \\
&~&-\frac{\epsilon}{\kappa \tau}\Bigg\{\mathcal{G}(\tau) - e^{-\kappa \tau}\mathcal{G}(0)\Bigg\},
\ee
where $W_{iso}^{T_h}$ is the work done during the corresponding isothermal process, i.e. $W_{iso}^{T_h}=T_h\log\big( Z(\delta_{min})/Z(\delta_{max})\big)$, with $Z$ being the partition function $Z: \delta \mapsto 1+e^{-\delta/T_h}$, $W_{ad}$ is the work done during the corresponding adiabatic process, i.e. $W_{ad} = \epsilon p_0$, where $p_0 = p(0)$ and $\epsilon =\delta_{max}-\delta_{min}$.
\end{theorem}
\bpr
We start by noting that 
\be\label{eq:lemma:<W>:1}
\frac{\D p}{\D \delta}&=&\frac{\D p}{\D t}.\frac{\D t}{\D \delta}= -\frac{\kappa \tau}{\epsilon}\Big(\gamma_h(\delta)-p\Big), 
\ee
where the last line follows from \eqref{eq:lemma:PTh_dp/dt} and \eqref{eq:PTh_ddelta/dt} while supressing the dependence on $t$. Integrating \eqref{eq:lemma:<W>:1} with respect to $\delta$ from $\delta_{max}$ to $\delta_{min}$, we have
\be\label{eq:lemma:<W>:2a}
\bigintssss_{\delta_{max}}^{\delta_{min}} p~\D \delta = \bigintssss_{\delta_{max}}^{\delta_{min}}\gamma_h\big(\delta\big)\D \delta +  \frac{\epsilon}{\kappa \tau}\bigintssss_{\delta_{max}}^{\delta_{min}} \frac{\D p}{\D \delta}\D \delta.
\ee
Then, plugging \eqref{eq:lemma:<W>:2a} in \eqref{eq:PTh_<W>_cts} implies
\be\label{eq:lemma:<W>:2b}
\mu_W(\tau)= -\bigintssss_{\delta_{max}}^{\delta_{min}}\gamma_h\big(\delta\big)\D \delta -  \frac{\epsilon}{\kappa \tau}\bigintssss_{\delta_{max}}^{\delta_{min}} \frac{\D p}{\D \delta}\D \delta.
\ee 
Substituting the expression for $\gamma_h(\delta)$ and evaluating the integral gives us the first term of \eqref{eq:lemma:<W>:2b} as
\be\label{eq:lemma:<W>:3}
\bigintssss_{\delta_{max}}^{\delta_{min}}\gamma_h\big(\delta\big) \D \delta = -T_h \ln\frac{Z(\delta_{min})}{Z(\delta_{max})},
\ee
where $Z$ is the partition function $Z : t \mapsto 1+e^{-\delta(t)/T_h}$. The expression above is simply the negative of the work done during the corresponding isothermal reversible process,
\be \label{eq:lemma:<W>:3a}
W_{iso}^{T_h}\triangleq T_h \ln\frac{Z(\delta_{min})}{Z(\delta_{max})}.
\ee
Next, we evaluate the integral in the second term in \eqref{eq:lemma:<W>:2b} using \cref{lemma:p(t)} together with the boundary conditions $p(\delta_{max})=p_0$ and $p(\delta_{min})=p(\tau)$. Thus, we have
\be\label{eq:lemma:<W>:4}
&~&\int_{\delta_{max}}^{\delta_{min}} \frac{\D p}{\D \delta}~\D \delta = p(\delta_{min})-p(\delta_{max})\notag \\
&=& p_0 \big(e^{-\kappa \tau} -1\big) + \mathcal{G}(\tau)-e^{-\kappa t}\mathcal{G}(0).
\ee
Now, if one changes the energy gap from $\delta_{max}$ to $\delta_{min}$ adiabatically the distribution of work is simply a two-point distribution, where $W=0$ occurs with probability $1-p_0$ and $W=\epsilon$ occurs with probability $p_0$. Thus, the average work done would be 
\be \label{eq:lemma:<W>:5}
W_{ad}\triangleq\epsilon p_0.
\ee
Plugging \eqref{eq:lemma:<W>:3} and \eqref{eq:lemma:<W>:4} in \eqref{eq:lemma:<W>:2b} together with \eqref{eq:lemma:<W>:3a} and \eqref{eq:lemma:<W>:5} gives us the result.
\epr 
\begin{corollary}
The expression for average work in \cref{theorem:average work} reduces to the adiabatic case in the limit $\tau\rightarrow 0$, i.e. 
\be \label{eq:lemma:<W>:W_ad}
 \lim_{\tau\rightarrow 0} \mu_W(\tau) = W_{ad},
\ee 
and the isothermal case in the limit $\tau\rightarrow\infty$,
\be \label{eq:lemma:<W>:W_iso}
  \lim_{\tau\rightarrow\infty} \mu_W(\tau) &=& W_{iso}^{T_h}.
\ee 
\end{corollary}
\bpr 
    Let us first derive the adiabatic limit, $\tau\rightarrow 0$:
    \be
    \lim_{\tau\rightarrow 0} \mu_W(\tau) &=&W_{iso}^{T_h} + \lim_{\tau\rightarrow 0}\Bigg\{\frac{W_{ad}}{\kappa \tau}\big(1-e^{-\kappa \tau}\big) + \notag \\
    &~&\frac{\epsilon}{\kappa \tau}\Bigg(\mathcal{G}(\tau) - e^{-\kappa \tau}\mathcal{G}(0)\Bigg)\Bigg\}.
    \ee
    Let us first look at the second term in the limit
    \be
    \lim_{\tau\rightarrow 0} \frac{1}{\tau}\big(1-e^{-\kappa \tau}\big) &=&  \lim_{\tau\rightarrow 0} \frac{1}{\tau} \Bigg(1-\bigg(1-\kappa \tau +\frac{\kappa^2 \tau^2}{2}-\cdots\bigg)\Bigg)\notag\\
    &=&\lim_{\tau\rightarrow 0} \Bigg(\kappa - \frac{\kappa ^2 \tau}{2}+\cdots\Bigg)=\kappa.
    \ee
    So, we have
    \be
  &~&\lim_{\tau\rightarrow 0} \mu_W(\tau) = W_{iso}^{T_h} + W_{ad} + \lim_{\tau\rightarrow 0}~\frac{\epsilon}{\kappa \tau}\Bigg\{\mathcal{G}(\tau)- e^{-\kappa \tau}\mathcal{G}(0)\Bigg\} \notag\\
    &=& W_{iso}^{T_h} + W_{ad} + T_h\Bigg\{-e^{-\frac{\delta_{min}}{T_h}}~\Phi_L\bigg(-e^{-\frac{\delta_{min}}{T_h}},1,1 \bigg) + \notag\\
    &~&e^{-\frac{\delta_{max}}{T_h}}~\Phi_L\bigg(-e^{-\frac{\delta_{max}}{T_h}},1,1 \bigg)\Bigg\},   
    \ee
where we have used \cref{def:Lerchi_Phi} in the second step. Since,
    $ 
  z~\Phi_L(z,1,1) = -\log (1-z),
    $ 
    we have 
    \be
    &~&\lim_{\tau\rightarrow 0} \mu_W(\tau) =W_{iso}^{T_h} + W_{ad} + \notag \\
    &~&T_h\Bigg\{\log{(1+e^{-\frac{\delta_{min}}{T_h}})} - \log{(1+e^{-\frac{\delta_{max}}{T_h}})}\Bigg\},
    \ee
    where the first term cancels the third term due to \eqref{eq:lemma:<W>:3}, and thus we obtain \eqref{eq:lemma:<W>:W_ad}. The isothermal limit, $\tau\rightarrow \infty$, can be similarly obtained since
    \be
  \lim_{\tau\rightarrow\infty}\mu_W(\tau)&=&W_{iso}^{T_h} - \lim_{\tau\rightarrow\infty}\Bigg\{- \frac{W_{ad}}{\kappa \tau}\big(1-e^{-\kappa \tau}\big) + \notag \\
   &~&\frac{\epsilon}{\kappa \tau}\Bigg(\mathcal{G}(\tau) -  e^{-\kappa \tau}\mathcal{G}(0)\Bigg)\Bigg\},
    \ee
  and it is clear that the second term in the equation above would vanish in the limit $\tau \rightarrow \infty$. Furthermore, using \cref{def:G}, we find that the third term would also vanish in the limit and so we recover \eqref{eq:lemma:<W>:W_iso}.
\epr
\subsection{Lower bound on variance}\label{subsec:lb-variance}
We will now establish by means of the following theorem that work fluctuations are non-zero for general partial thermalization processes independent of \cref{assumption:1}.

\begin{theorem}[Fluctuations in work]\label{lemma:fluctuations}

Consider a two-level system undergoing a finite-time process such that the energy gap is driven from $\delta_{max}$ to $\delta_{min}$ in $L$ discrete steps along with partial thermalizations for a finite time $\tau$. Then, the following are true in general for the random variable $W_L$ denoting the total work done during the process:
\be\label{eq:lemma:Fluctuations:1}
&~&\lim_{L\rightarrow\infty}\Pr[W_L=0]=\big(1-p_0\big)e^{-\kappa\int_{0}^\tau \D t \gamma_h(\delta(t))},
\ee
and 
\be\label{eq:lemma:Fluctuations:2}
&~&\lim_{L\rightarrow\infty}\Pr[W_L=\epsilon]=p_0 e^{-\kappa \int_0^\tau \D t(1-\gamma_h(\delta(t))},
\ee
where $\epsilon = \delta_{max}-\delta_{min}$, $L\rightarrow\infty$ is the continuous time limit, and $p_0$ is the initial excited state probability of the two-level system.
\end{theorem}
\bpr 
From \cref{tab:table6}, it is easy to see that the following expression holds for a discrete partial thermalization process composed of $L$ steps such that each step takes time $\Delta t$:
\be
\Pr[W_L=0]&=&\big(1-p_0\big)\prod_{l=1}^{L-1}\Big(1-\kappa\Delta t\gamma_h\big(\delta(l\Delta t)\big)\Big).
\ee
Taking $\log$ on both-sides of the above equation we have,
\be
\log\Big(\Pr[W_L=0]\Big)&=&\log\big(1-p_0\big)+\notag \\
&~&\sum_{l=1}^{L-1}\log\Big(1-\kappa\Delta t\gamma_h\big(\delta(l \Delta t)\big)\Big)\notag\\
&\mysimeqDeltat&~\log\big(1-p_0\big)-\notag \\
&~&\kappa\sum_{l=1}^{L-1}\Delta t\gamma_h\big(\delta(l \Delta t)\big).
\ee
Taking the limit $\Delta t\rightarrow 0$ ($L\rightarrow\infty$) and observing that the second term above would thus be a Riemann sum, we obtain \eqref{eq:lemma:Fluctuations:1} by exponentiating the resulting expression (and noting that limit commutes with continuous functions). Similarily, the last row in \cref{tab:table6} implies that 
\be
&&\Pr[W_L = \epsilon]=p_0\prod_{l=1}^{L-1}\Big(1-\kappa\Delta t\big(1-\gamma_h\big(\delta(l\Delta t)\big)\big)\Big).
\ee
Again, taking $\log$ on both-sides we have
\be\label{eq:lemma:Fluctuations:6}
&~&\log\Big(\Pr[W_L= \epsilon]\Big)\notag \\
&=&\log p_0+\sum_{l=1}^{L-1}\log\Big(1-\kappa\Delta t\Big(1-\gamma_h\big(\delta(l \Delta t)\big)\Big)\Big)\notag\\
&\mysimeqDeltat&~\log p_0-\kappa\sum_{l=1}^{L-1} \Delta t\Big(1-\gamma_h\big(\delta(l \Delta t)\big)\Big).
\ee
Again, taking the limit $\Delta t\rightarrow 0$ ($L\rightarrow\infty$) results in an expression that gives \eqref{eq:lemma:Fluctuations:2} upon exponentiation.
\epr 
This result analytically establishes that the distribution of work is typically broad as was also found numerically in Ref.~\onlinecite{Dhar_PRE2005}. While the theorem above holds in general, a lower bound on the variance of work done by systems driven linearly in time (\cref{assumption:1}) can be obtained as a corollary to it.
\begin{corollary}[Lower bound on variance of work]\label{corollary-lb-variance}
For a finite-time process as per \cref{assumption:1} along with partial thermalizations, the variance of work is bounded from below as
\be \label{eq:corollary:variance_lb}
&~&\sigma^2_W (\tau) \geq \big(1-p_0\big)\Bigg(\frac{Z(\delta_{min})}{Z(\delta_{max})}\Bigg)^{-\frac{\kappa \tau T_h}{\epsilon}}\mu_W^2(\tau) + \notag \\
&~&p_0~e^{-\kappa \tau}\Bigg(\frac{Z(\delta_{min})}{Z(\delta_{max})}\Bigg)^{\frac{\kappa \tau T_h}{\epsilon}}\Big(\epsilon + \mu_W(\tau)\Big)^2,
\ee 
where $Z$ is the partition function $Z:\delta\mapsto1+e^{-\delta/T_h}$ and $\mu_W(\tau)$ is the average work output of the process as given by \cref{theorem:average work}. Moreover, the lower bound is saturated in the adiabatic limit,
\be \label{eq:corollary:variance_lb:ad}
\lim_{\tau\rightarrow 0}\sigma^2_W(\tau) = p_0 (1-p_0)\epsilon^2,
\ee  
as well as in the isothermal limit, 
\be \label{eq:corollary:variance_lb:iso}
\lim_{\tau\rightarrow\infty}\sigma^2_W(\tau)=0.
\ee
\end{corollary}
\bpr 
We will first derive expressions for the probabilities of work values $W=0$ and $W=\epsilon$ when undergoing a finite-time process s per \cref{assumption:1} along with partial thermalizations using \cref{lemma:fluctuations}. Using the expression for $\delta(t)$ as given by \eqref{eq:lemma:p(t):3} we change the variable of integration to $\delta$ in \eqref{eq:lemma:Fluctuations:1} and obtain the following after taking log on both sides:
\be
&~&\lim_{L\rightarrow\infty}\log\Big(\Pr[W_L=0]\Big)\notag \\
&=&\log\Big(1-p_0\Big) + \frac{\kappa \tau}{\epsilon}\bigintssss_{\delta_{max}}^{\delta_{min}}\frac{\D \delta}{1+e^{\delta/T_h}}.
\ee
Evaluating the integral and exponentiating the above we have
\be\label{eq:Pr[W=0]}
\lim_{L\rightarrow\infty}\Pr[W_L=0]&=&~\big(1-p_0\big)\Bigg(\frac{Z(\delta_{min})}{Z(\delta_{max})}\Bigg)^{-\frac{\kappa \tau T_h}{\epsilon}},
\ee
where $Z$ is the partition function $Z:\delta\mapsto 1+e^{-\delta/T_h}$. Similarly,  \eqref{eq:lemma:Fluctuations:2} gives 
\be
&~&\lim_{L\rightarrow\infty}\log\Big(\Pr[W_L= \epsilon]\Big)\notag\\
&=&\log p_0 + \frac{\kappa \tau}{\epsilon} \bigintssss_{\delta_{max}}^{\delta_{min}} \frac{\D \delta}{1+e^{-\delta/T_h}}.
\ee
Again, evaluating the integral and exponentiating the above we obtain
\be \label{eq:Pr[W=epsilon]}
\lim_{L\rightarrow\infty}\Pr[W_L=\epsilon]&=&p_0~e^{-\kappa \tau}\Bigg(\frac{Z(\delta_{min})}{Z(\delta_{max})}\Bigg)^{\frac{\kappa T T_h}{\epsilon}}.
\ee 
Now that we have the expressions for $\Pr[W_L=0]$ and $\Pr[W_L=\epsilon]$ it is straightforward to obtain a lower bound for the variance of work as the sum of these two contributions. Thus,
\be
\sigma^2_{W_L}(\tau) &\geq & \Pr[W_L=0]\big(\mu_{W_L}(\tau)\big)^2 + \notag \\
&~&\Pr[W_L= \epsilon ]\big(\epsilon - \mu_{W_L}(\tau)\big)^2.
\ee  
Taking the limit $L\rightarrow\infty$ and assuming that $W_L$ converges in probability to the random variable $W$ for the continuous process, we have 
\be
\sigma^2_{W}(\tau) &\geq & \lim_{L\rightarrow\infty}\Bigg\{\Pr[W_L=0]\big(\mu_{W}(\tau)\big)^2 + \notag \\
&~&\Pr[W_L= \epsilon ]\big(\epsilon - \mu_{W}(\tau)\big)^2\Bigg\}.
\ee  
Plugging \eqref{eq:Pr[W=0]} and \eqref{eq:Pr[W=epsilon]} in the equation above gives \eqref{eq:corollary:variance_lb}. Let us now look at the lower bound in the following two limiting cases.
\begin{itemize}
    \item Adiabatic limit, $\tau\rightarrow 0$:
    	\be \label{eq:corollary:ad_limit1}
    		&~&\lim_{\tau\rightarrow 0}\Bigg\{\big(1-p_0\big)\Bigg(\frac{Z(\delta_{min})}{Z(\delta_{max})}\Bigg)^{-\frac{\kappa \tau T_h}{\epsilon}}\mu_W^2(\tau) + \notag \\
    		&~&p_0~e^{-\kappa \tau}\Bigg(\frac{Z(\delta_{min})}{Z(\delta_{max})}\Bigg)^{\frac{\kappa \tau T_h}{\epsilon}}\Big(\epsilon - \mu_W(\tau)\Big)^2\Bigg\} \notag \\
    		&=& \lim_{\tau\rightarrow 0}\Bigg\{\big(1-p_0\big)\mu_W^2(\tau) + p_0\Big(\epsilon - \mu_W(\tau)\Big)^2\Bigg\} \notag\\
    		&=&\big(1-p_0\big)\epsilon^2p_0^2 + p_0\Big(\epsilon - \epsilon p_0\Big)^2 \notag \\
    		&=& p_0 (1-p_0)\epsilon^2, 
    	\ee 
   where the last line follows from \eqref{eq:lemma:<W>:W_ad}. Recall that the average work done when changing the energy gap from $\delta_{max}$ to $\delta_{min}$ adiabatically is given by \eqref{eq:lemma:<W>:5}. Moreover, the variance of work for an adiabatic process can be obtained by noting that the distribution of $W_{ad}\in\{0,\epsilon\}$ is simply $\{1-p_0,p_0\}$, i.e.
   \be \label{eq:corollary:ad_limit2}
   \sigma_W^2(\tau=0) &=& p_0 (\epsilon -W_{ad})^2 + (1-p_0)W_{ad}^2 \notag \\
   &=& p_0 (1-p_0)\epsilon^2.
   \ee   
  Therefore, \eqref{eq:corollary:ad_limit1} and \eqref{eq:corollary:ad_limit2} together imply that the lower bound is saturated in the said limit.
    \item Isothermal limit, $\tau\rightarrow\infty$:
\be\label{eq:corollary:iso_limit1}
&~&\lim_{\tau\rightarrow \infty}\Bigg\{\big(1-p_0\big)\Bigg(\frac{Z(\delta_{min})}{Z(\delta_{max})}\Bigg)^{-\frac{\kappa \tau T_h}{\epsilon}}\mu_W^2(\tau) + \notag \\
		&~&p_0~e^{-\kappa \tau}\Bigg(\frac{Z(\delta_{min})}{Z(\delta_{max})}\Bigg)^{\frac{\kappa \tau T_h}{\epsilon}}\Big(\epsilon - \mu_W(\tau)\Big)^2\Bigg\}.
\ee  

    Now, let us look at the relevant part in the first term of \eqref{eq:corollary:iso_limit1}. Plugging in the definition for the partition function $Z$, we have 
    \be
    	\lim_{\tau\rightarrow \infty}\Bigg(\frac{Z(\delta_{min})}{Z(\delta_{max})}\Bigg)^{-\frac{\kappa \tau T_h}{\epsilon}}\notag \\
    	=\lim_{\tau\rightarrow \infty}\Bigg(\frac{1+e^{-\delta_{min}/T_h}}{1+e^{-\delta_{max}/T_h}}\Bigg)^{-\frac{\kappa \tau T_h}{\epsilon}}&=& 0,
    \ee 
    as $1+e^{-\delta_{min}/T_h}>1+e^{-\delta_{max}/T_h}$. Similarly, we look at the relevant part of the second term in \eqref{eq:corollary:iso_limit1} to obtain
    \be
    	&~&\lim_{\tau\rightarrow \infty}e^{-\kappa \tau}\Bigg(\frac{Z(\delta_{min})}{Z(\delta_{max})}\Bigg)^{\frac{\kappa \tau T_h}{\epsilon}}\notag\\ 
    	&=&\lim_{\tau\rightarrow \infty}\Bigg(e^{-\epsilon/T_h}\frac{Z(\delta_{min})}{Z(\delta_{max})}\Bigg)^{\frac{\kappa \tau T_h}{\epsilon}} \notag\\
    	&=& \lim_{\tau\rightarrow \infty}\Bigg(e^{-\epsilon/T_h}\Bigg(\frac{1+e^{-\delta_{min}/T_h}}{1+e^{-\delta_{max}/T_h}}\Bigg)\Bigg)^{\frac{\kappa \tau T_h}{\epsilon}} \notag\\
    	&=& \lim_{\tau\rightarrow \infty}\Bigg(\frac{1+e^{\delta_{min}/T_h}}{1+e^{\delta_{max}/T_h}}\Bigg)^{\frac{\kappa \tau T_h}{\epsilon}}\notag\\
    	&=& 0,
    \ee 
where we have used the fact that $\epsilon = \delta_{max} - \delta_{min}$ along with $1+e^{\delta_{min}/T_h}<1+e^{\delta_{max}/T_h}$. Thus, we obtain     
	 \be\label{eq:corollary:iso_limit2}
    		\lim_{\tau\rightarrow \infty}\sigma^2_W(\tau) \geq 0.
     \ee 		
       Moreover, from Ref.~\onlinecite{Aberg-work-extraction_2013} we know that isothermal work extraction is fluctuation-free, i.e. 
   \be \label{eq:corollary:iso_limit3}
   	\sigma_W^2(\tau = \infty) =0.
   \ee 
  Again, \eqref{eq:corollary:iso_limit2} and \eqref{eq:corollary:iso_limit3} together  imply that the lower bound is saturated in this limit.
\end{itemize}
\epr 
\section{Numerical results}\label{sec:numerical_results}
In this section we present the results of the Monte Carlo simulation for the Markov process, \cref{fig:Markovchain}, to obtain estimates of the variance as a function of the time period of the process. Furthermore, for a two-level system that is initially in equilibrium with the bath, we find that the variance can be estimated using Jarzynski's fluctuation-dissipation relation.
\subsection{Monte Carlo for variance of work}
\begin{figure}[t] 
    \includegraphics[width=\columnwidth]{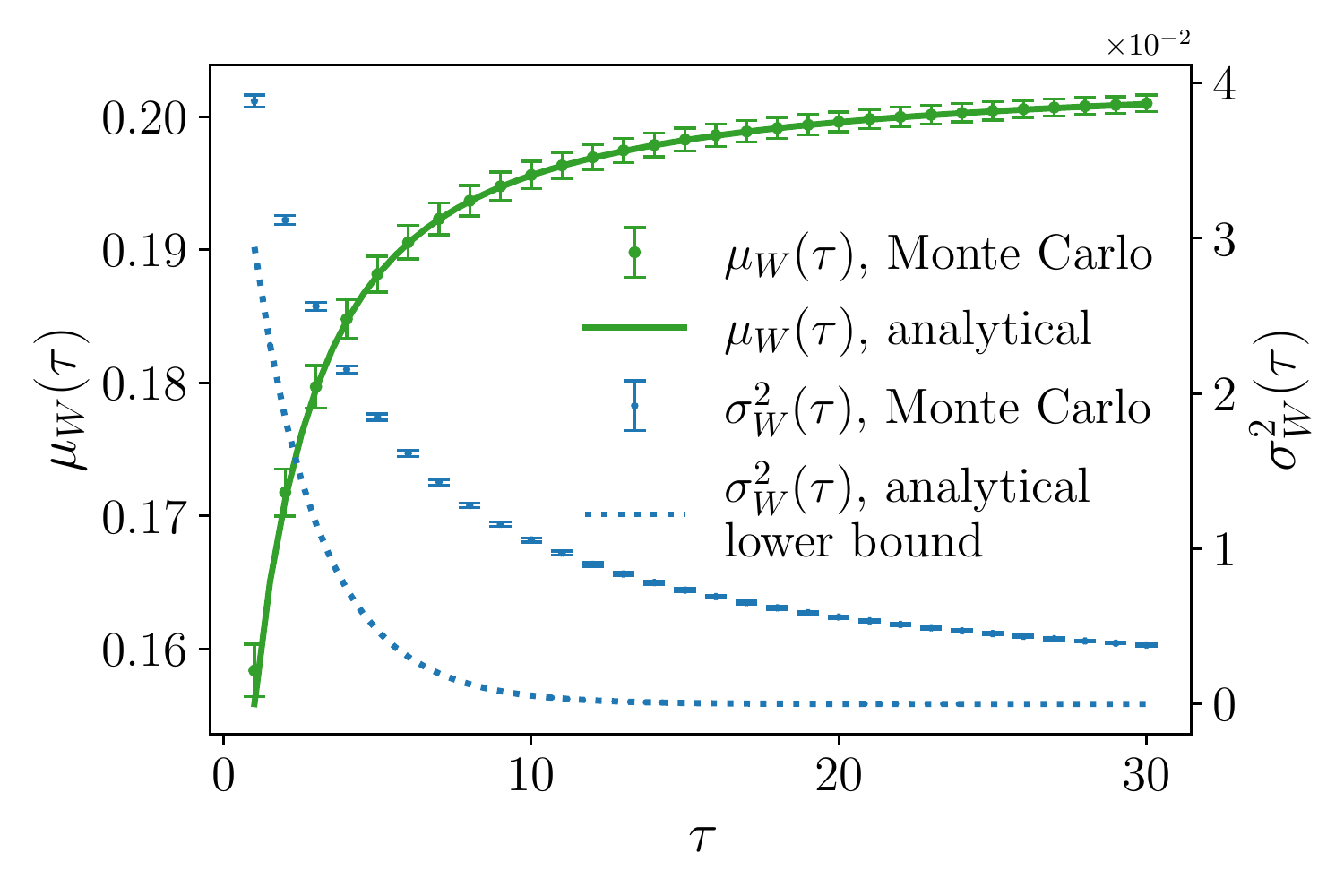}
    \caption{(Color online) On the x-axis we have total time of the work extraction process, $\tau$. The solid green (gray) curve interpolates between the adiabatic $W_{ad}=0.134$ and the isothermal $W_{iso}^{T_h}=0.204$ limits. The solid blue (black) curve also interpolates between the adiabatic $\sigma_W^2(\tau =0) = $ and the isothermal $\sigma_W^2(\tau = \infty)=0$ limits. The Monte Carlo simulations were done with $L=1000$ steps, where $L$ is the discretization (see \cref{eg:1}) and for integer values of $\tau\in [1, 30]$. The parameter values used are $\delta_{max}=1$, $\delta_{min}=0.5$, $T_h=2$, and $p_0 = \frac{1}{1+e}$.}
    \label{fig:W_avg-W_var}
\end{figure}

In order to compare the gap between the analytical lower bound obtained in \cref{corollary-lb-variance} with the actual variance, we perform Monte Carlo simulations since an analytical derivation seems to be intractable  owing to the time-dependent nature of the Markov process, \cref{fig:Markovchain}. The Monte Carlo basically simulates a discrete version of the Markov process under \cref{assumption:1}, see \cref{eg:1}. We plot the results of the same in \cref{fig:W_avg-W_var}. As a test of credibility, we find that the error-bars on our numerically obtained values of average work successfully envelop the analytical form as a function of $\tau$ (\cref{theorem:average work}). The error bars were obtained using $10^4$ independent runs. The independent runs were parallelized using GNU parallel \cite{gnu_parallel}.
\begin{figure}[t]
    \includegraphics[width=\columnwidth]{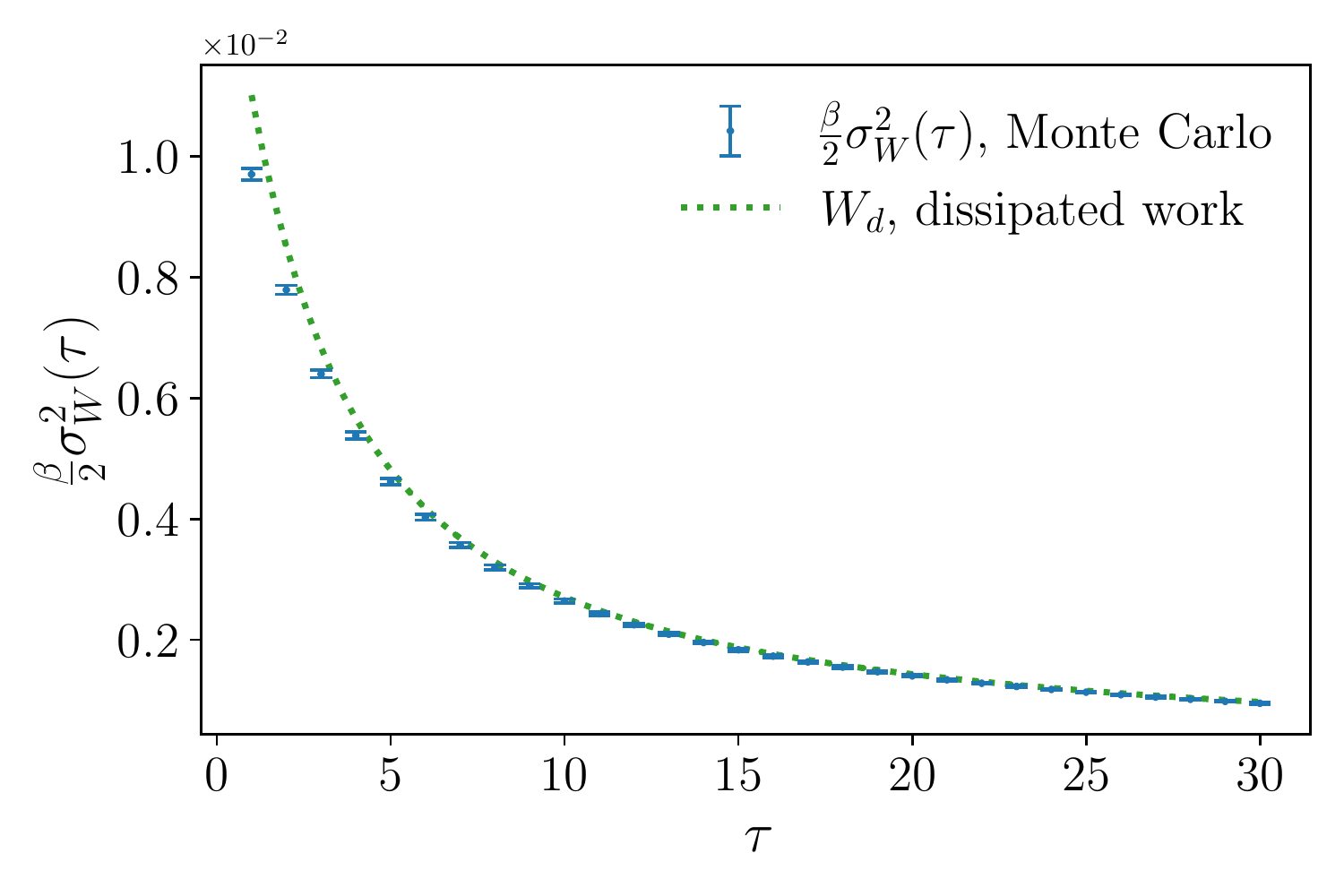}
    \caption{(Color online) Fluctuation-dissipation relation for finite-time processes. On the x-axis we have the total time of the work extraction process, $\tau$. $W_d$ is the dissipated work. The Monte Carlo simulations were done for $L=1000$ steps, where $L$ is the discretization (see \cref{eg:1}) and for integer values of $\tau\in [1, 30]$. The parameter values used are $\delta_{max}=1$, $\delta_{min}=0.5$, $T_h=2$, and $p_0 = \frac{1}{1+e^{1/2}}$. }
    \label{fig:fluc-diss}
\end{figure}
\subsection{Fluctuation-dissipation relation}
A fluctuation-dissipation relation governing an irreversible thermodynamic process is a statement about the relation between the dissipated work (on average) when a system is driven away from equilibrium and the corresponding fluctuations of work during such a process. Jarzynski's \cite{Jarzynski_PRL1997} much touted result gave such a relation in the weak system-bath interaction limit. Basically, once the system is in equilibrium with the ambient bath it is disconnected from the bath and then the work extraction process is performed which essentially amounts to changing the value of some relevant parameter (that governs of the Hamiltonian) over a finite amount of time. When the time over which the process is carried out---the switching time---is large enough it renders the distribution of work Gaussian.  Denoting the random variable for the work done during such an irreversible process by $W$ and its mean and variance by $\mu_W$ and $\sigma^2_W$ respectively, the dissipated work is $W_{diss}= \mu_W - \Delta F$, the difference between the average work done during the process, $\mu_W$, and the average work done during the corresponding reversible process i.e.~the free-energy difference, $\Delta F$. The fluctuation-dissipation relation can then be expressed as
\be\label{eq:fluc-diss}
W_{diss} = \frac{\beta}{2}\sigma^2_W,
\ee 
where $\beta=1/k_B T_h$ with $T_h$ being the temperature of the ambient bath. Using \cref{theorem:average work} with \mbox{$p_0=1/(1+e^{\delta_{max}/T_h})$}, we plot the dissipated work, $\mu_W(\tau) - \Delta F$, and the estimate of the variance from the Monte Carlo simulation as a function of the total time period of the process $\tau$ in \cref{fig:fluc-diss}. We find that dissipated work provides an upper bound for the variance of work in general. This bound is saturated in the limit of large $\tau$.
\section{Application to finite-time heat engines}\label{sec:application}

In this section we discuss finite-time heat engines operating in cycles that are composed of work extraction processes involving partial thermalizations and instantaneous adiabatic energy-level transformations.

Finite-time heat engines are characterized by their non-zero power output in contrast to the ideal Carnot engine. In this section we first review the Carnot engine for a two-level system and then study one such engine that incorporates work extraction processes mediated by partial thermalizations, replacing the ideal isothermal processes of the Carnot cycle. We optimize the power output of such cycles for fixed time periods over different set of parameters and constraints in \cref{subsubsec:1} and \cref{subsubsec:2}. Finally, we compare the two in \cref{subsec:compare-finite-time-optimal-cycles}. 
\subsection{Carnot engine: review}
\begin{figure}[h]
	\includegraphics[scale=0.45]{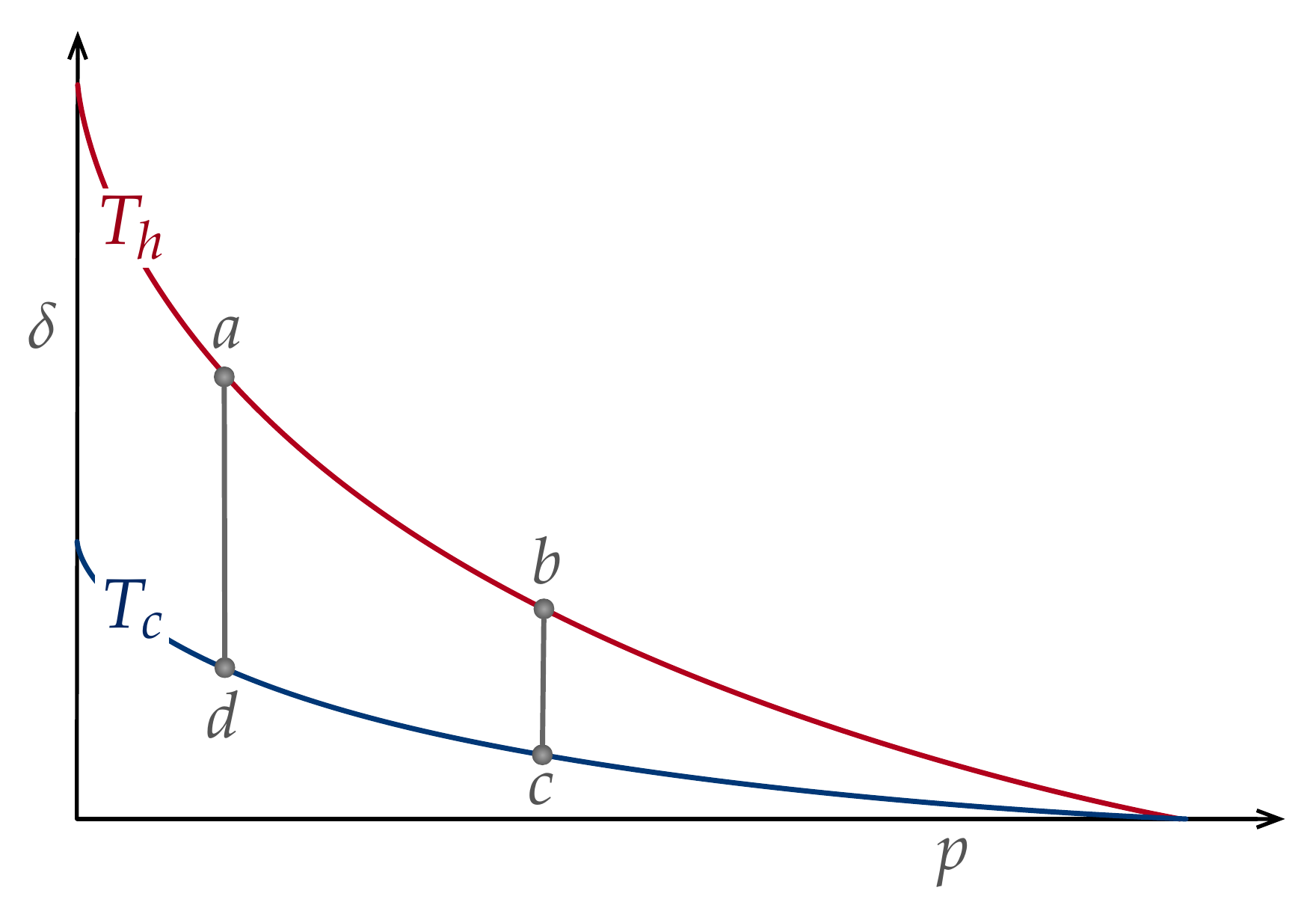}
	\caption{Carnot cycle for a two-level system.}
	\label{fig:CarnotCycle}	
\end{figure}

Let us assume that we have access to a hot bath at temperature $T_h$, a cold bath at temperature $T_c$, and a two-level system whose energy gap $\delta$ can be varied over a fixed range between $\delta_{min}$ and $\delta_{max}$. As done in the previous section, let us set the ground state energy of the system to be zero. Then, a quantum Carnot cycle \cite{Quan_qthermo-cycles-I2007} which is composed of four stages, like it's classical counterpart, can be defined using points $a, b, c, d$ on the $p-\delta$ plot for a two-level system, \cref{fig:CarnotCycle}. First, we have
\begin{itemize}[leftmargin=*]
    \item $a\mapsto b$, an isothermal expansion: at point $a$, the system is in a Gibbs-thermal state at temperature $T_h$ with an energy gap $\delta_a$. The occupation probability for the excited state is $p_a=\frac{1}{1+e^{\delta_a/T_h}}$. During a quantum isothermal expansion, the energy levels must change such that all energy levels are scaled by the same factor $k_1<1$, see \cref{app:2} for proof. The state of the qubit at point $b$ is a Gibbs state at temperature $T_h$ with an energy gap $\delta_b = k_1\delta_a$. As has been shown in Ref.~\onlinecite{Aberg-work-extraction_2013} the work done during such a reversible isothermal process is essentially \emph{deterministic}, and is given by    \be\label{W_ab}
    		 W_{ab}=T_h\ln{\frac{Z(\delta_b)}{Z(\delta_a)}},
    	\ee
    	where $Z$ is the partition function $Z: \delta \mapsto 1+e^{-\delta/T_h}$.
       \item Then, $b\mapsto c$, an adiabatic process: during a general quantum adiabatic process the energy levels of the system change without any accompanying change in occupation probabilities and thus there is no change in entropy. In particular, at this stage the energy levels are changed by a factor such the system is in the Gibbs state with respect to the cold bath at temperature $T_c$ implying 
       	\be \label{eq:carnot_cycle:1}
       		\frac{\delta_b}{T_h} = \frac{\delta_c}{T_c}.
       	\ee 
       		
       Thus, the energy gap of the system at point $c$ is $\delta_c = T_h/T_c~k_1\delta_a$. The work done in this process $W_{bc}$ is a \emph{random variable} as it depends on the state of the system at point $b$\footnote{This means that if the system was in the ground state then it continues to be in the ground state of the new Hamiltonian.}. The average value of the work done during this process is 
    \be 
     \overline{W}_{bc}=\frac{1}{1+e^{\delta_c/T_c}}(\delta_b-\delta_c).
    \ee 
    \item Next, the compression stage with $c\mapsto d$, an isothermal compression: again, during this process the energy levels are scaled by a factor $k_2 >1$. So, the system is still in a Gibbs-thermal state with respect to the cold bath at temperature $T_c$ at point $d$ but with an energy gap $\delta_d = T_h/T_c~k_1k_2\delta_a$. The work cost of this process is deterministic and is given by 
    \be\label{W_bc}
     W_{cd}=-T_c\ln{\frac{Z(\delta_d)}{Z(\delta_c)}}.
     \ee 

    \item Finally, we have $d\mapsto a$, an adiabatic process where the energy gap is changed such that we go back to the starting point $a$ with energy gap $\delta_a$ such that  the excited state occupation probability $p_d=p_a$. Therefore,
    \be \label{eq:carnot_cycle:2}
    \frac{\delta_d}{T_c}=\frac{\delta_a}{T_h}.
    \ee
    But, $\delta_d = T_h/T_c~k_1k_2\delta_a$. This implies that the constant $k_2$ is not independent but must satisfy the following relation $k_2=\frac{1}{k_1}$. And, the average work cost of this process is 
    \be
    \overline{W}_{da}=\frac{1}{1+e^{\delta_a/T_h}}(\delta_a-\delta_d).
    \ee 
\end{itemize}
The total work done during the Carnot cycle described above, denoted by the random variable $W_C$, is given by
\be
W_C= W_{ab}+ W_{bc} - W_{cd} - W_{da}.
\ee
\begin{lemma}\label{lemma:carnot-cycle}
 The total work done $W_C$ during a microscopic implementation of the Carnot cycle is a random variable distributed according to a four-point distribution listed in the table below.

\begin{table}[H]
  \begin{center}
  	
    \label{tab:table2}
    \begin{tabular}{|c|c|c|c|c|c|c|c|} 
      ~~~~$W_C=W_{ab}+W_{bc}-W_{cd}-W_{da}$~~~~ & ~~~~$\Pr{[W_C]}$~~~~\\
      \hline
      \\[-1em]
      $(T_h-T_c)\ln{Z(\delta_b)/Z(\delta_a)}+$\textcolor{RubineRed}{0} & $(1-p_a)(1-p_b)$\\
      \hline
      \\[-1em]
      $(T_h-T_c)\ln{Z(\delta_b)/Z(\delta_a)}$\textcolor{ForestGreen}{$-(\delta_a-\delta_d)$} & $(1-p_b)p_a$\\
      \hline
      \\[-1em]
     $(T_h-T_c)\ln{Z(\delta_b)/Z(\delta_a)}+$\textcolor{Cyan}{$\delta_b-\delta_c$} & $(1-p_a)p_b$\\
      \hline
      \\[-1em]
      $(T_h-T_c)\ln{Z(\delta_b)/Z(\delta_a)}+$\textcolor{BlueViolet}{$\delta_b-\delta_c-\delta_a+\delta_d$} & $p_bp_a$
     \end{tabular}
  \end{center}
\end{table}
\noindent
The expected efficiency of the Carnot cycle is 
\be \label{eq:eff_C}
\eta_C^{avg} = \bigg(1 -\frac{T_c}{T_h}\bigg).
\ee  
\end{lemma}
\bpr 
$W_{ab}$ and $W_{cd}$ are essentially deterministic and are given by  \eqref{W_ab} and \eqref{W_bc} while $W_{bc}$ and $W_{da}$ are random variables. In \cref{tab:table3}, we list all the possible states that the system could be in at each of the four nodes \textit{a, b, c, d} and thus obtain all the possible values for $W_{bc}-W_{da}$. 
\begin{table*}[t]
    \begin{tabular}{|c|c|c|c|c|c|c|c|} 
      ~~~~$a\mapsto$~~~~ & ~~~~$b\mapsto$~~~~ & ~~~~$c\mapsto$~~~~ & ~~~~$d\mapsto$~~~~  & ~~~~$a$~~~~ & ~~~~$W_{bc}-W_{da}$~~~~ & ~~~~$\Pr{[W_{bc}-W_{da}]}$~~~~\\
      \hline
      0 & 0 & 0 & 0 & 0 & \textcolor{RubineRed}{0} & $(1-p_a)^2(1-p_b)$\\
      \hline
      0 & 0 & 0 & 1 & 1 & \textcolor{ForestGreen}{$-(\delta_a-\delta_d)$} & $(1-p_a)(1-p_b)p_a$\\
      \hline
      0 & 1 & 1 & 0 & 0 & \textcolor{Cyan}{$\delta_b-\delta_c$} & $(1-p_a)^2p_b$\\
      \hline
      0 & 1 & 1 & 1 & 1 & \textcolor{BlueViolet}{$\delta_b-\delta_c -\delta_a+\delta_d$} & $(1-p_a)p_bp_a$ \\
      \hline
      1 & 0 & 0 & 0 & 0 & \textcolor{RubineRed}{0} & $p_a(1-p_a)(1-p_b)$\\
      \hline
      1 & 0 & 0 & 1 & 1 & \textcolor{ForestGreen}{$-(\delta_a-\delta_d)$} & $p_a^2(1-p_b)$\\
      \hline
      1 & 1 & 1 & 0 & 0 & \textcolor{Cyan}{$\delta_b-\delta_c$} & $p_ap_b(1-p_a)$\\
      \hline
      1 & 1 & 1 & 1 & 1 & \textcolor{BlueViolet}{$\delta_b-\delta_c -\delta_a+\delta_d$} & $p_a^2p_b$
     \end{tabular}
     \caption{Occupation of the ground state is designated by $0$ and that of the excited state by $1$. The Carnot cycle is $a\mapsto b\mapsto  c\mapsto d\mapsto a$. Starting at $a$ with the system in the ground state as one completes the cycle the system could be in the excited state---it undergoes thermalization from $c\mapsto d$. Same colour entries under the $W_{bc}+W_{da}$ column are identical and the probabilities corresponding to those entries add up.}
      \label{tab:table3}
\end{table*}   
Thus, we can obtain an expression for the average work done by simply multiplying and adding the corresponding entries of columns $W_C$ and $\Pr{[W_C]}$ to arrive at
\be \label{eq:avg_work_C}
\mu_{W_C} &=& \big(T_h-T_c\big)\ln\frac{Z(\delta_b)}{Z(\delta_a)} - \Bigg(1 - \frac{T_c}{T_h}\Bigg)\delta_a p_a + \notag \\
&~& \Bigg(1-\frac{T_c}{T_h}\Bigg)\delta_b p_b \notag\\
&=& \big(T_h-T_c\big)\ln\frac{Z(\delta_b)}{Z(\delta_a)} - \Bigg(1 - \frac{T_c}{T_h}\Bigg)\big(\delta_a p_a - \delta_b p_b\big)\notag\\
&=& \Bigg(1 - \frac{T_c}{T_h}\Bigg)\Bigg(T_h \ln\frac{Z(\delta_b)}{Z(\delta_a)} + \delta_b p_b - \delta_a p_a\Bigg).
\ee 
We can then derive the average Carnot efficiency $\eta_C^{avg}$ by dividing the average work done by the heat input (which is when the system undergoes isothermal expansion from point $a$ to $b$), $Q_{ab}$. The heat exchanged during a process, denoted by $Q$, is given by the first law of Thermodynamics, i.e. $ Q=\Delta U-W$, where $\Delta U$ is the change in the total energy of the system during the process while $W$ is the corresponding work yield/cost. In our case, the process is an isothermal expansion $a\mapsto b$, so $\Delta U_{ab}= \big(\delta_b p_b - \delta_a p_a\big)$ and $W=-W_{ab}$ where $W_{ab}$ is the deterministic work yield of the process and is given by \eqref{W_ab}. Thus, we have 
\be\label{eq:carnot_cycle:4}
Q_{ab} =  \big(\delta_b p_b - \delta_a p_a\big) + T_h \ln\frac{Z(\delta_b)}{Z(\delta_a)}.
\ee 
Since efficiency is defined as the ratio of the work output and the heat input, \eqref{eq:avg_work_C} and \eqref{eq:carnot_cycle:4} imply \eqref{eq:eff_C}.
\epr 
Now, for the given pair of temperatures $T_c$ and $T_h$ the Carnot efficiency is the  maximum attainable efficiency. It is independent of the points $a,~b,~c,$ and $d$ on the $p-\delta$ plot, \cref{fig:CarnotCycle}, that define a work extraction cycle for the engine connecting the two isotherms. However, there is another quantity that becomes relevant under the constraint of being able to vary the energy gap $\delta$ only between $\delta_{min}$ and $\delta_{max}$. The cycle that maximizes the average work output is the Carnot cycle that encloses the largest area on the $p-\delta$ plot---it maximizes both efficiency and average work. We state this intuition in the lemma below deferring a formal proof to \cref{app:3} for completeness.
\begin{lemma}[Optimal Carnot cycle for average work]\label{lemma:Carnot_cycle_optimal}\label{lemma:opt-Carnot}
	Given a Carnot engine formed by a two-level system operating between a hot bath at temperature $T_h$ and a cold bath at temperature $T_c$ such that the energy gap of the system $\delta$ can only be varied over a fixed range between $\delta_{\min}$ and $\delta_{\max}$, the cycle (defined by the points $a,~b,~c,$ and $d$ on the $p-\delta$ plot) that maximizes the average work output of the Carnot engine is the one for which $\delta_a = \delta_{\max}$ and $\delta_c = \delta_{\min}$.
\end{lemma}

 The power output of such a cycle is zero due to the isothermal processes that require infinitely long equilibration times. But, finite-time work extraction cycles have non-zero power output and for such cycles one is generally interested in the efficiency at maximum power \cite{Review_Stochastic-thermodynamics-fluctuation-theorems-molecular-machines_Seifert_2012}. We analyze such engines in subsequent section.

 \subsection{Finite-time heat engines}
 For constant time periods, maximizing power amounts to maximizing the average work output. We define a modification of the Carnot cycle that incorporates the finite-time element---replacing isothermal processes in a Carnot cycle by work extraction processes with partial thermalizations. So, a finite-time cycle denoted by $\mathpzc{a}\mapsto \mathpzc{b} \mapsto \mathpzc{c} \mapsto \mathpzc{d} \mapsto \mathpzc{a}$ on the $p-\delta$ plot constitutes a sequence of four processes. First we have
\begin{itemize}[leftmargin=*]
	\item $\mathpzc{a}\mapsto \mathpzc{b}$, work extraction with partial thermalizations with respect to the hot bath. The coordinates of point $\mathpzc{a}$ on the $p-\delta$ plot are ($\delta_{\mathpzc{a}},~p_{\mathpzc{a}}$). The system is driven under \cref{assumption:1} by an amount $\delta_a - \delta_{b1}$ for a time $\tau_1$. The occupation probability for the excited state $p_{\mathpzc{b}}(\tau_1)$ can then be obtained using \cref{lemma:p(t)}. Furthermore, the average work done during this process would be given by \cref{theorem:average work}. 
	\item Then, $\mathpzc{b} \mapsto \mathpzc{c}$, an adiabatic process. The energy gap is changed from $\delta_{\mathpzc{b}}$ to $\delta_{\mathpzc{c}}$ keeping the occupation probabilities fixed, i.e. $p_{\mathpzc{c}} = p_{\mathpzc{b}}(\tau_1)$. 
	The average work done during this process would be 
		\be \label{eq:finite_time_cycle:1}
			\overline{W}_{\mathpzc{b}\mathpzc{c}} = p_{\mathpzc{b}}(\tau_1)\big(\delta_{\mathpzc{c}} - \delta_{\mathpzc{b}}\big).
		\ee 
	\item Next, we have $\mathpzc{c} \mapsto \mathpzc{d}$, work extraction with partial thermalizations with respect to the cold bath. Starting from the point $\mathpzc{c}$ with coordinates ($\delta_{\mathpzc{c}},~p_{\mathpzc{b}}(\tau_1)$) the system is again driven under \cref{assumption:1} for a time $\tau_2$ such that the energy gap increases from $\delta_{\mathpzc{c}}$ to $\delta_{\mathpzc{d}}$. To ensure that we complete the cycle and reach point $\mathpzc{a}$ in the end $\delta_{\mathpzc{d}}$ must be such that 
		\be \label{eq:finite_time_cycle:2}
			p_{\mathpzc{d}}(\tau_2) = p_{\mathpzc{a}}.
		\ee    
		An expression for $p_{d_1}(\tau_2)$ and average work cost of this process can be derived along the lines of \cref{lemma:p(t)} and \cref{theorem:average work} as done in \cref{app:4}.
		
	\item Finally, we close the loop with $\mathpzc{d} \mapsto \mathpzc{a}$ adiabatically. Having reached $\delta_{\mathpzc{d}}$ in accordance with \eqref{eq:finite_time_cycle:2}, we complete the cycle by changing the energy gap keeping the occupation probabilities fixed. The average work cost of this process is simply
		\be 
			\overline{W}_{\mathpzc{d}\mathpzc{a}} = p_{\mathpzc{a}} \big(\delta_{\mathpzc{a}} -\delta_{\mathpzc{d}}\big).
		\ee 
\end{itemize} 

The time period of the cycle as described above would thus be $\mathcal{T} = \tau_1 + \tau_2$. Since we are interested in the efficiency at maximum power, we want to maximize the average work output of a finite-time cycle with a fixed time period $\mathcal{T} = \tau_1 + \tau_2$, which would simply be the sum of the average work done at each of the four steps described above. The parameters characterising a finite-time cycle as described above are given by the set $\{\delta_{\mathpzc{a}},~p_{\mathpzc{a}}, ~\delta_{\mathpzc{b}}, ~\delta_{\mathpzc{c}}, ~\tau_1\}$. As $\mathcal{T} =\tau_1 + \tau_2$, there is only one free parameter---we choose it to be $\tau_1$. The fact that for every value of $\tau_2$ one has to solve \eqref{eq:finite_time_cycle:2} for $\delta_{\mathpzc{d}}$ leaves no room for analytical analysis. We perform numerical optimizations instead. We first consider the special case where one can recover the Carnot cycle, \cref{lemma:opt-Carnot}, in the limit of large $\mathcal{T}$. The numerical optimizations were performed on \emph{Mathematica} \cite{Mathematica} using the \emph{Nelder-Mead} method \cite{Mathematica-ref}.
 \subsubsection{Optimal finite-time cycles limiting to Carnot cycle}\label{subsubsec:1}
In order to recover the Carnot cycle in the limit of large time period of a finite-time cycle, we need to fix the values of the parameters accordingly. For the first process to approach the hot isotherm, it is clear that $\delta_{\mathpzc{a}}$ and $\delta_{\mathpzc{b}}$ should be the same as in the case of the optimal Carnot cycle, \cref{lemma:opt-Carnot}. However, $\delta_{\mathpzc{c}}$ must be chosen to lie on the cold isotherm, i.e. it should satisfy the relation
\be 
p_{\mathpzc{b}}(\tau_1) = \frac{1}{1 + e^{\delta_{\mathpzc{c}}/T_c}},
\ee 
\begin{figure}[t]
\includegraphics[width=\columnwidth]{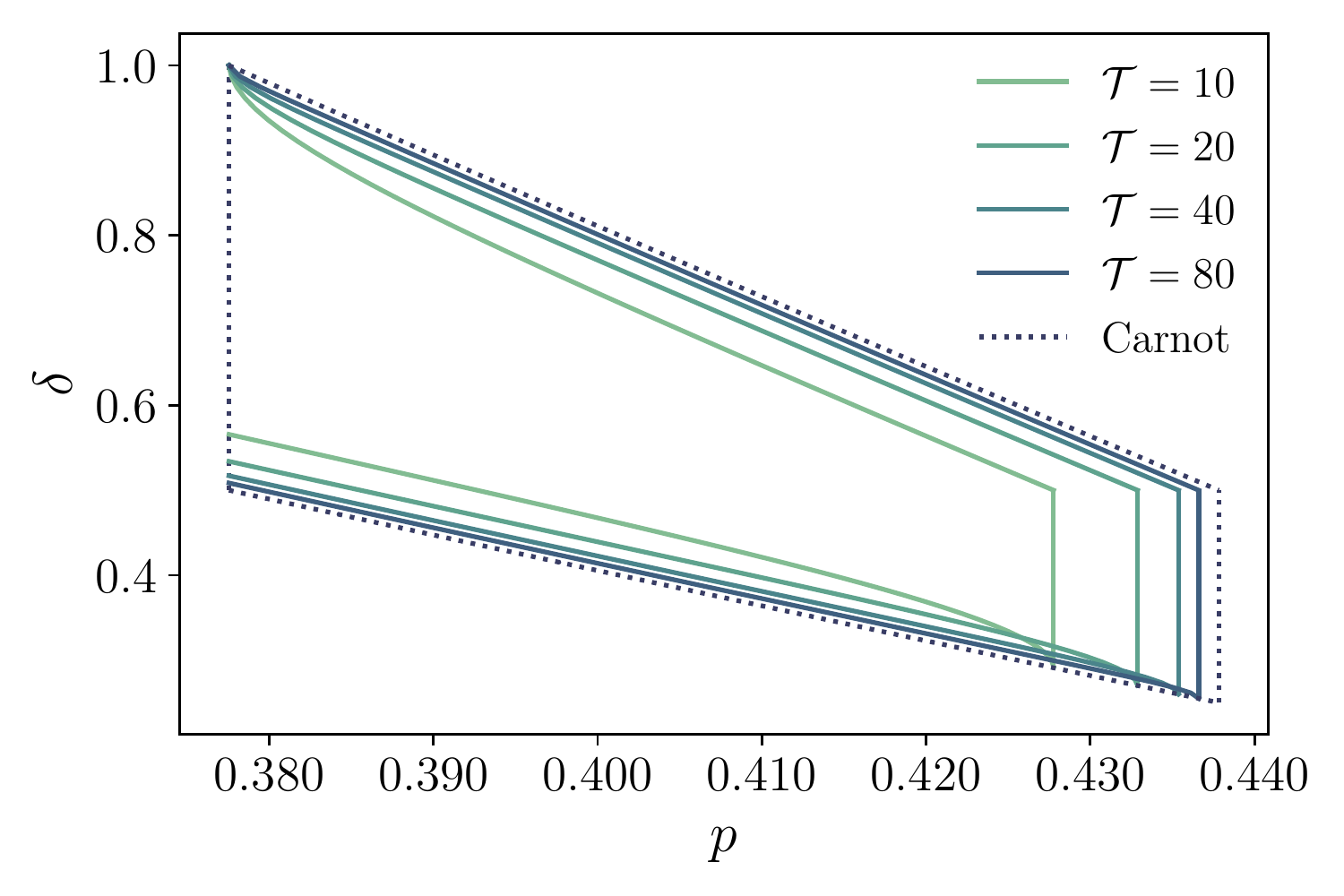}
 \caption{(Color online) On the x-axis we have occupation probability for the excited state $p$. On the y-axis we have the energy gap $\delta$. We use parameters $\delta_{max}=1$, $\delta_{min}=0.25$, $T_h=2$, $T_c=1$, $\delta_a = \delta_{\max}$, and $\delta_b = 2\delta_{\min}$.}
\label{fig:opt-cycles-1-param}
\end{figure}
\begin{figure}[t]
\includegraphics[width=\columnwidth]{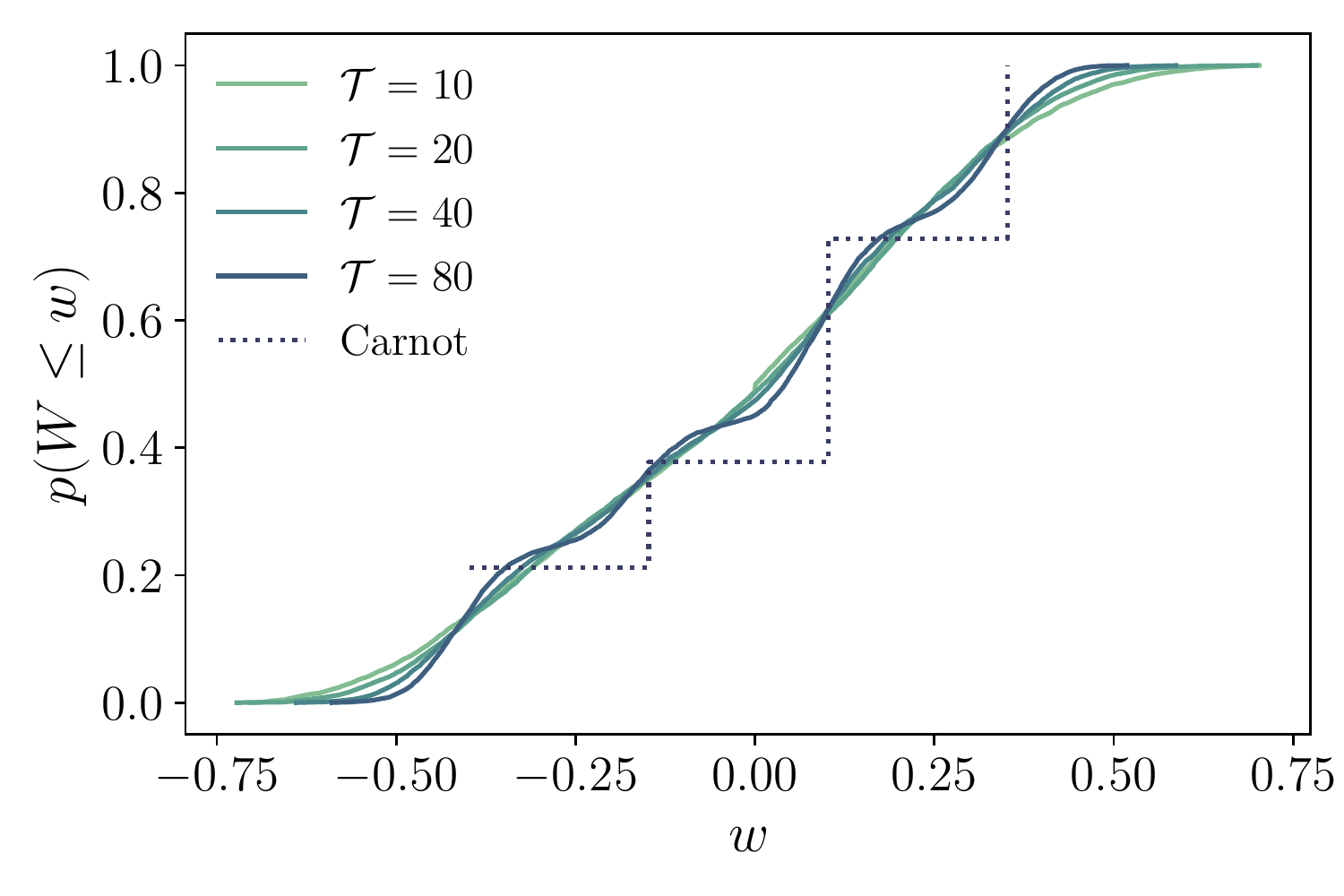}
\caption{(Color online) On the y-axis we have the cumulative distribution function for the random variable W, the work extracted during different optimal cycles (different values of $\mathcal{T}$). On the x-axis we have the possible work values. The Monte Carlo simulations were performed for $10^4$ samples for each of the optimum cycles.}
\label{fig:opt-cycles-cdf-1-param}
\end{figure}
\begin{figure}[t]
\includegraphics[width=\columnwidth]{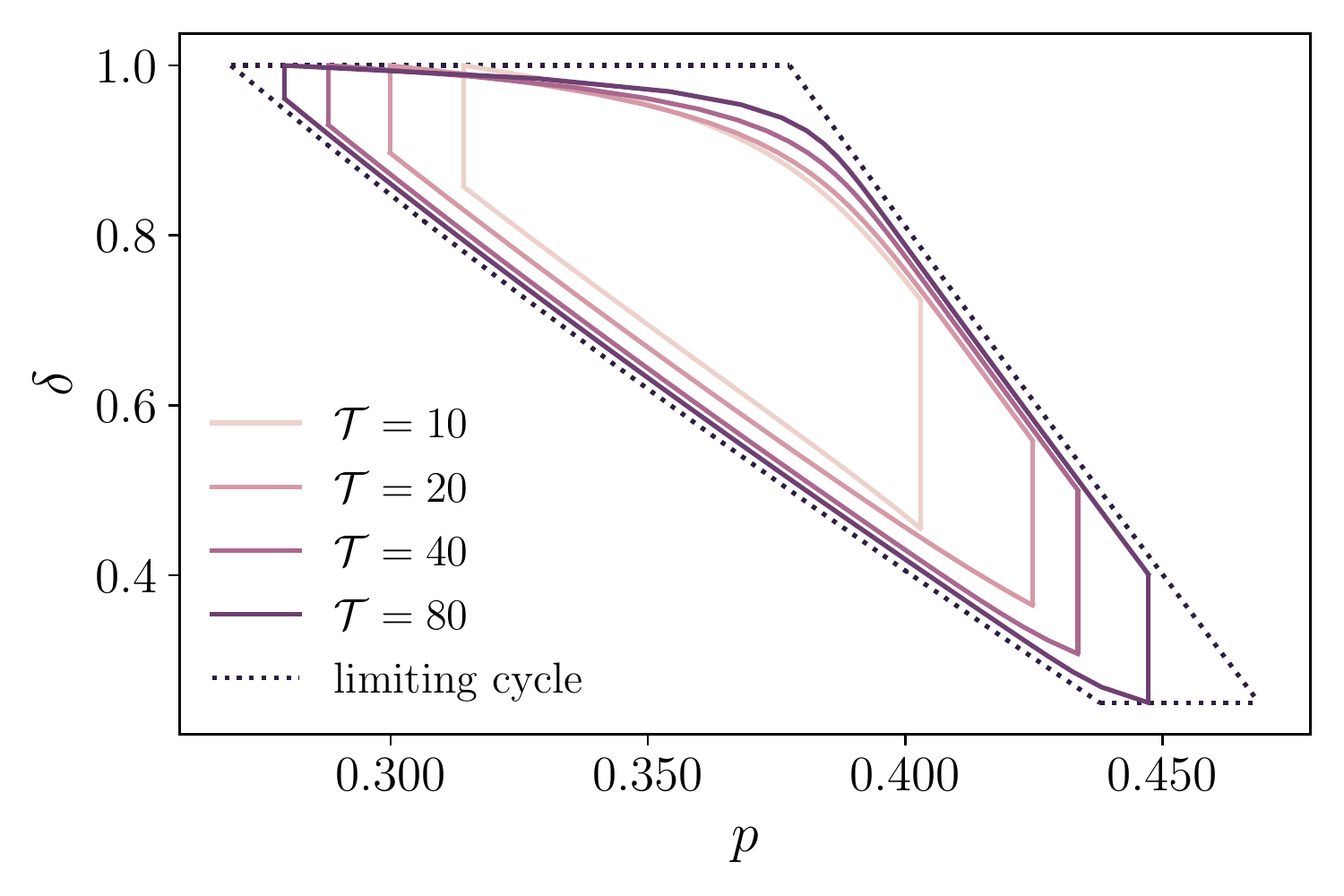}
\caption{(Color online) On the x-axis we have occupation probability for the excited state $p$. On the y-axis we have the energy gap $\delta$. We use parameters $\delta_{max}=1,~\delta_{min}=0.25,~T_h=2, ~T_c=1.$}
\label{fig:opt-cycles-general}
	\end{figure}
	\begin{figure}[t]
\includegraphics[width=\columnwidth]{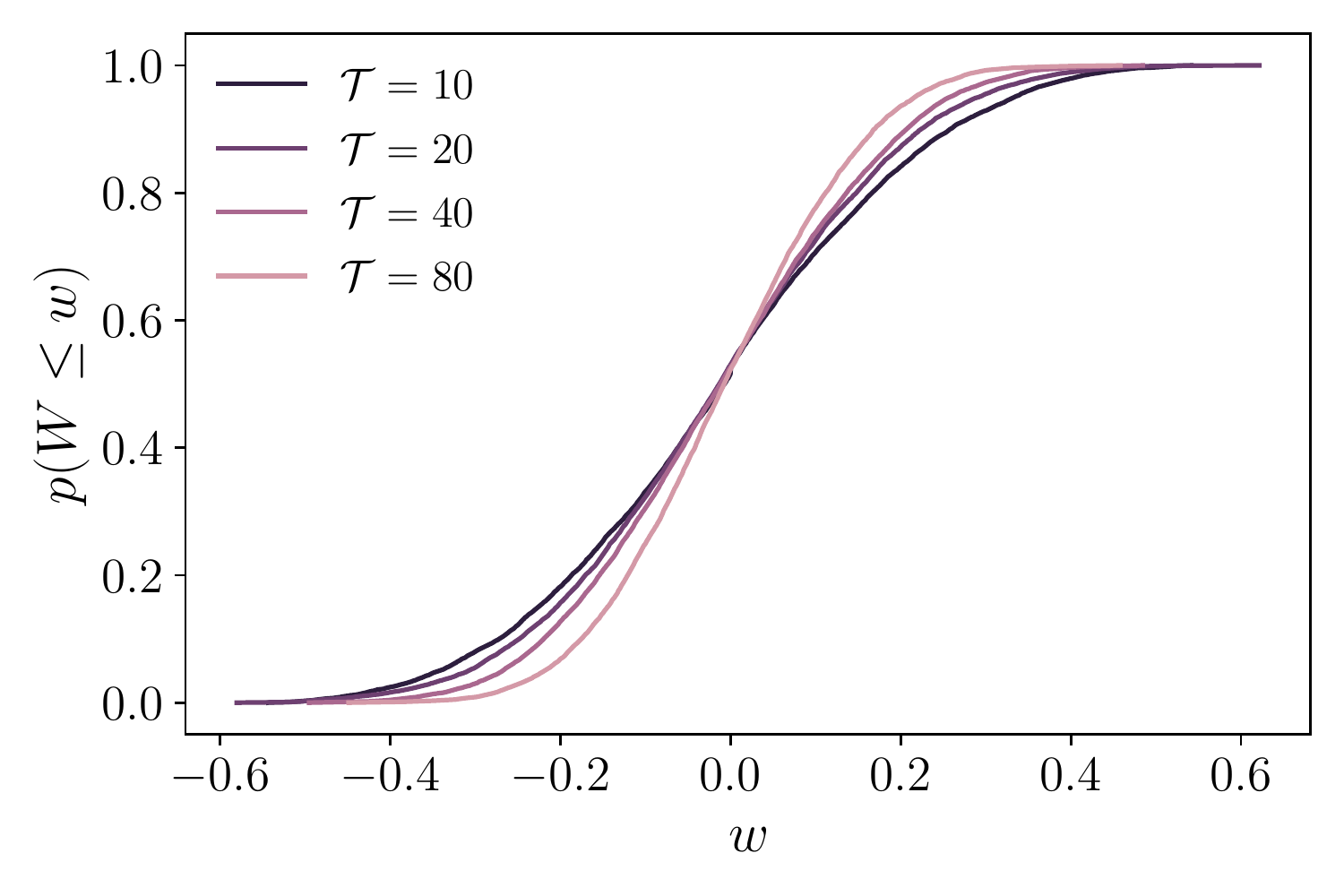}
\caption{(Color online) On the y-axis we have the cumulative distribution function for the random variable W, the work extracted during different optimal cycles (different values of $\mathcal{T}$). On the x-axis we have the possible work values. The Monte Carlo simulations were performed for $10^4$ samples for each of the optimum cycles.}
\label{fig:opt-cycles-cdf-5-params}
\end{figure}

since only then would the third leg, i.e. $\mathpzc{c} \mapsto \mathpzc{d}$ approach the cold isotherm in the limit of large $\mathcal{T}$. This implies that we are left with only one free parameter, namely $\tau_1$. Then, maximizing average work output for different values of  
$\mathcal{T}$ results in different optimal cycles which we plot in \cref{fig:opt-cycles-1-param}. We also plot the cumulative distribution for the different optimal cycles along with that of the Carnot cycle to study the fluctuations as we approach equilibrium in \cref{fig:opt-cycles-cdf-1-param}. Note that the Carnot cycle has a four-point work distribution, see \cref{lemma:carnot-cycle}. The distributions for finite-time cycles are obtained by performing Monte Carlo simulations. We find that even though the average work cycles start approaching the Carnot cycle quickly the cumulative distribution still remains smooth until we go to very large values of $\mathcal{T}$.    
     



\subsubsection{General optimal finite-time cycles}\label {subsubsec:2}
Previously we were interested in the special case of the power maximization problem that gave the Carnot cycle in the limit of large time periods. However, for the most general problem, where one has access to a hot bath at temperatures $T_h$ and a cold one at temperature $T_c$ and the energy gap can only be driven between $\delta_{\max}$ and $\delta_{\min}$, one should optimize all the parameters in the set  $\{\delta_{\mathpzc{a}},~p_{\mathpzc{a}}, ~\delta_{\mathpzc{b}}, ~\delta_{\mathpzc{c}}, ~\tau_1\}$. Here, we find that the optimal cycle in the limit of large time period approaches a different cycle; one where the two isotherms are connected by two purely thermal processes. So, $\mathpzc{d}\mapsto\mathpzc{a}$ and $\mathpzc{b}\mapsto\mathpzc{c}$ would be thermalizations connecting the two isotherms at $\mathpzc{d}=\mathpzc{a}=\delta_{\max}$ and $\delta_{\mathpzc{b}}=\delta_{\mathpzc{c}}=\delta_{\min}$ respectively as shown in \cref{fig:opt-cycles-cdf-compare}. This can be understood intuitively as we want to maximize work \emph{output}---the processes where we have to perform work are not favourable.  Since work output of an adiabatic process is less than the corresponding isothermal process and vice-versa for work input, the adiabatic legs are completely lost and get replaced by isothermal extensions. Even though this cycle is not very relevant from the point of view of power maximization since for large time periods power is no longer a meaningful metric, it is worth noting the curious form of the cyclic process in contrast to the maximum efficiency cycle---the Carnot cycle. We plot the optimal cycles for different values of time periods $\mathcal{T}$ in \cref{fig:opt-cycles-general} along with the cumulative distributions in \cref{fig:opt-cycles-cdf-5-params}. 
   
 \subsection{Comparing finite-time optimal cycles}\label {subsec:compare-finite-time-optimal-cycles}
\begin{figure}[t]
\includegraphics[width=\columnwidth]{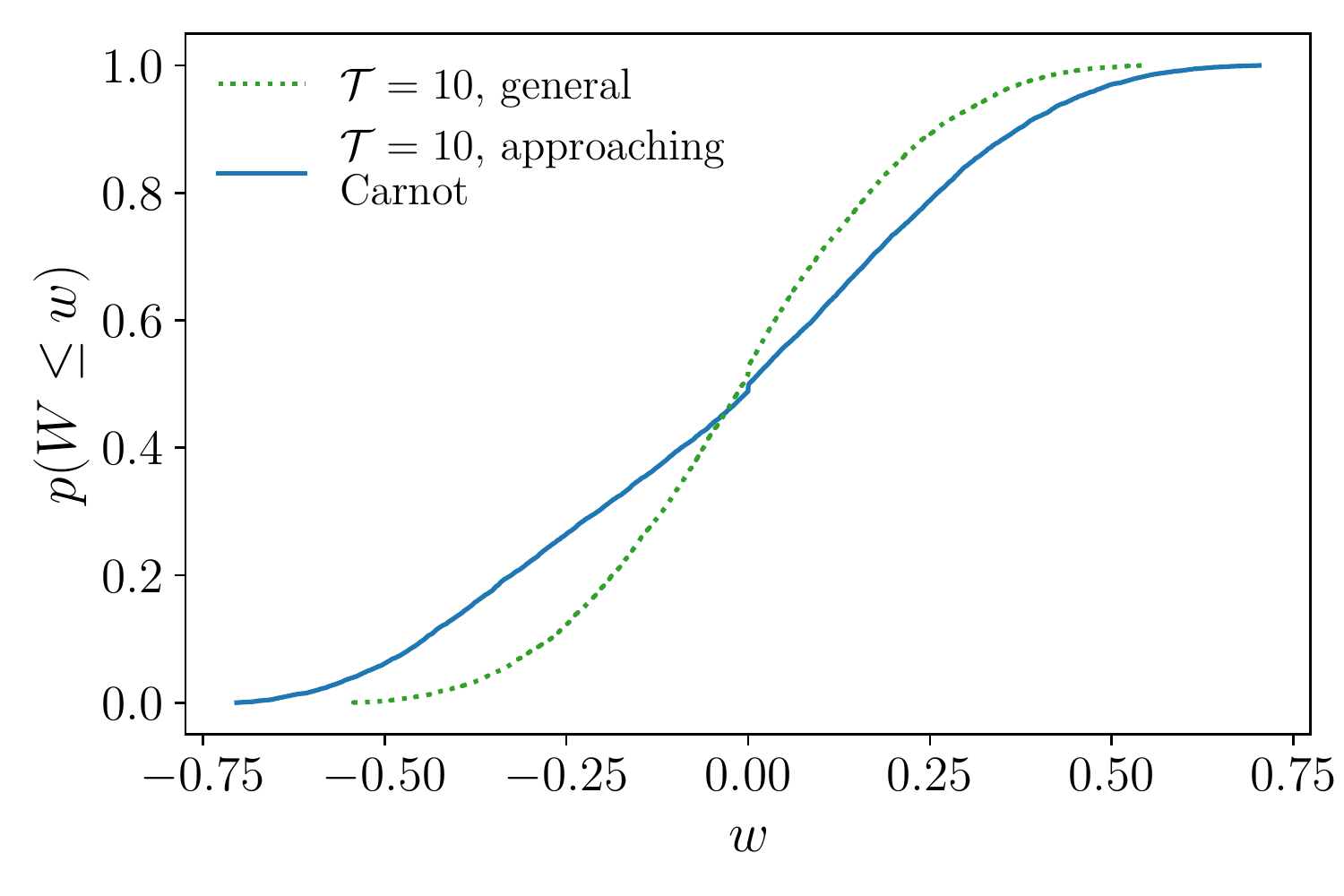}
\caption{(Color online) Comparing cumulative distribution of work for different types of optimal cycles. On the x-axis we have the possible work values. Negative values of $w$ imply a net work input.}
\label{fig:opt-cycles-cdf-compare}
\end{figure}

\begin{figure}[t]
\includegraphics[width=\columnwidth]{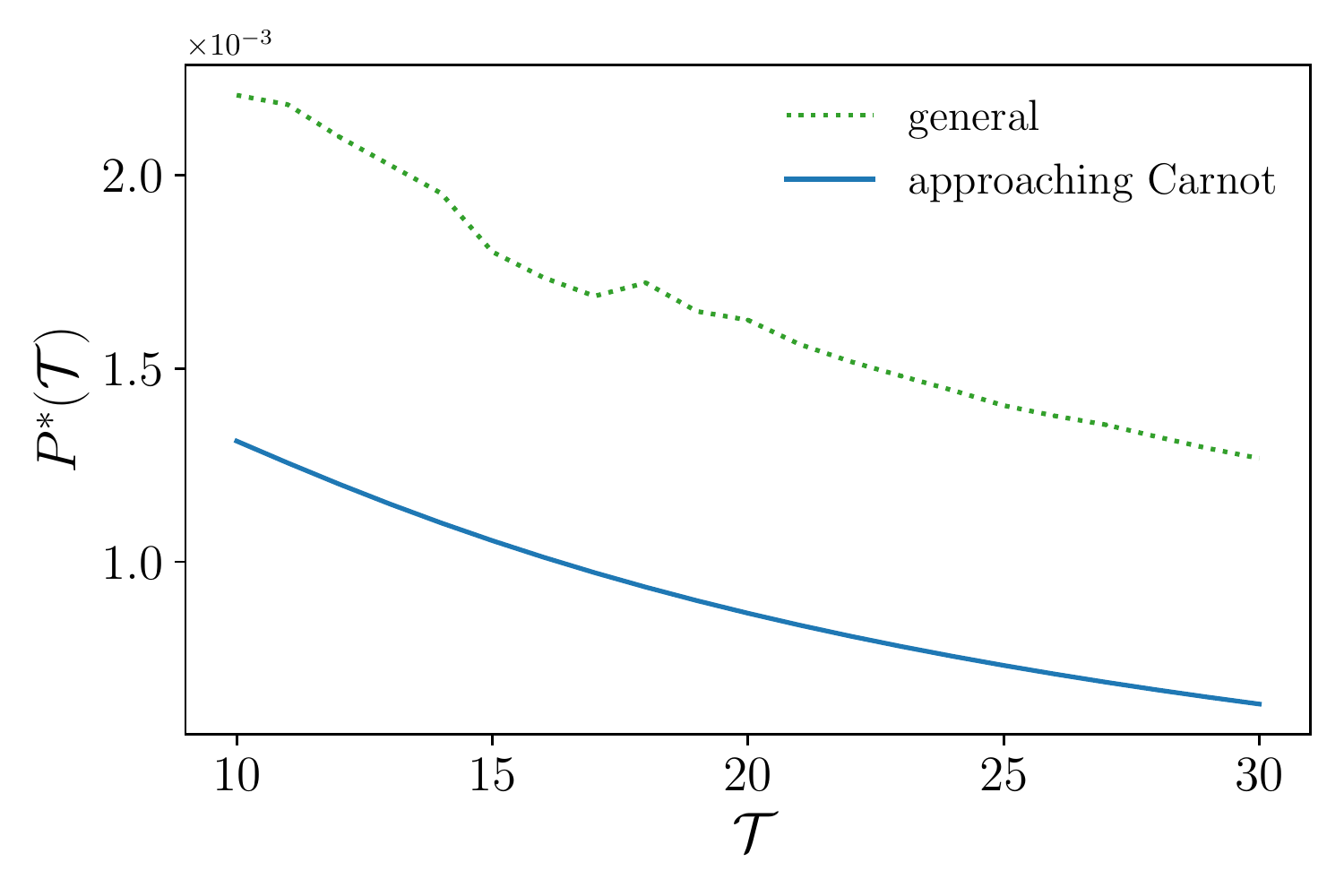}
\caption{(Color online) Comparing maximum power output for different type of cycles. On the y-axis we have the efficiency at maximum power $P^*(\mathcal{T})$. On the x-axis we have the time period of the finite-time cycles, $\mathcal{T}$. We use parameters $T_h=2$, $T_c=1$, $\eta_C=0.5$.}
\label{fig:max-power-compare}
\end{figure}

We compare the two scenarios discussed above in terms of their cumulative distributions and find that the general optimal cycles have a better quality of work---less fluctuations. For example, in \cref{fig:opt-cycles-cdf-compare} we plot the distributions for $\mathcal{T}=10$ and observe that the cumulative distribution for the solution of the general optimum problem crosses the one obtained in \cref{subsubsec:1} around $w=0$ and lies below it for almost all negative values of $w$. This means that the probability with which one has to input work in the former case is always less than the latter. Intuitively, there is no real reason to constrain the parameter values as we did in \cref{subsubsec:1} other than the imposed restriction of recovering the Carnot cycle in the limit of large $\mathcal{T}$. This limit is not particularly interesting from the point of view of maximizing power as it vanishes in the said limit. However, such a comparison is at the level of fluctuations only. Next, we compare $P^*$, the maximum power itself as a function of $\mathcal{T}$ for the two cases in \cref{fig:max-power-compare} and find that the general optimal power is higher than the corresponding power from optimal cycles that approach the Carnot cycle in the limit of large time periods. This is what one would expect anyway as the latter is a restricted version of the general optimization problem, \cref{subsubsec:2}.
\begin{figure}[t]
\includegraphics[width=\columnwidth]{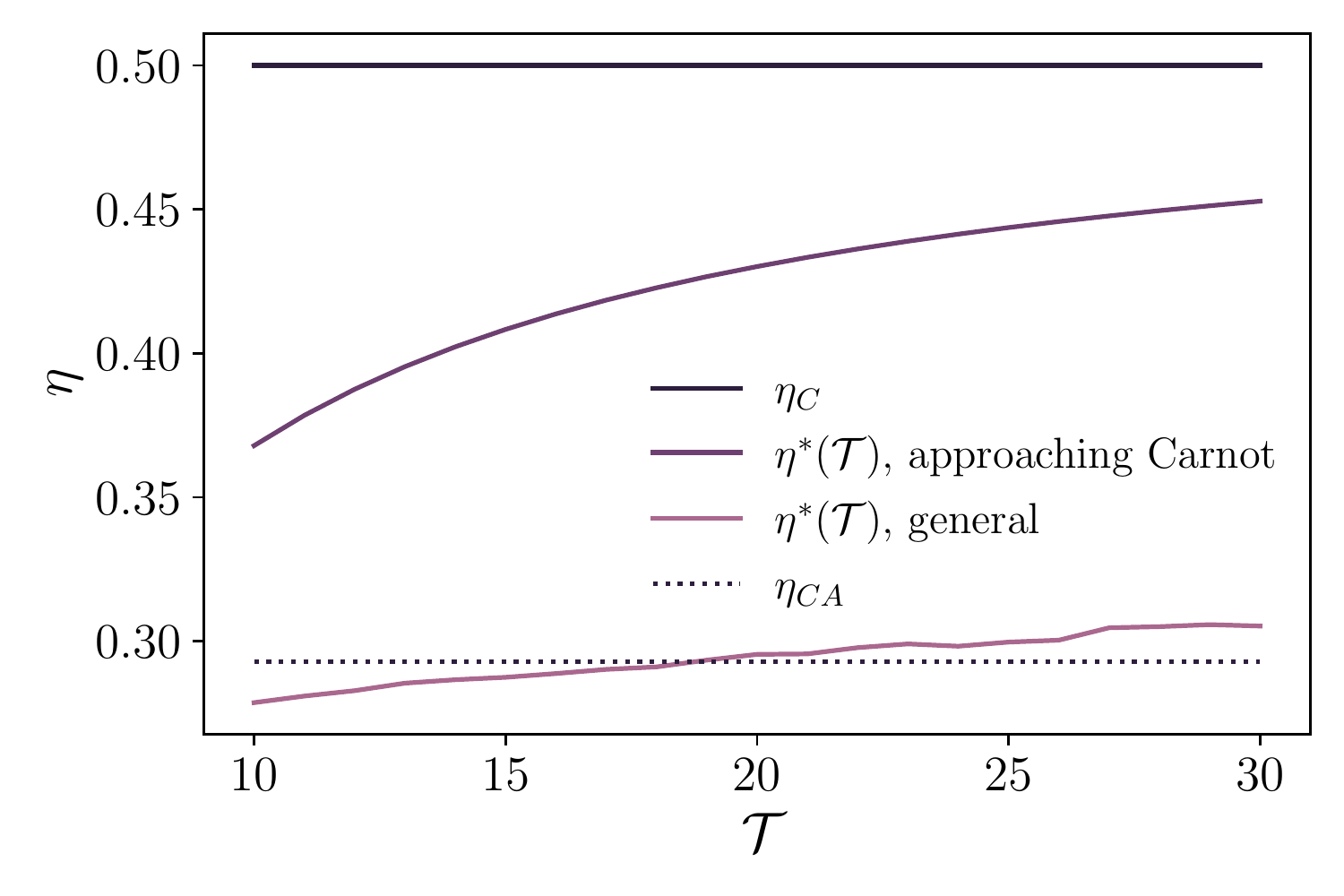}
\caption{(Color online) Comparing efficiency at maximum power for different types of cycles. On the x-axis we have the time period of the finite-time cycles, $\mathcal{T}$. On the y-axis we have efficiency $\eta$. We use parameters $T_h=2$, $T_c=1$, $\eta_C=0.5$, and $\eta_{CA}=0.293$.}
\label{fig:eff-max-power-compare}
\end{figure}
\cref{fig:opt-cycles-cdf-compare} and \cref{fig:max-power-compare} together imply that the general optimal cycles are better as far as power output and fluctuations are concerned. Finally, we compare the optimal efficiencies $\eta^*$ for the two scenarios as a function of $\mathcal{T}$ in \cref{fig:eff-max-power-compare}. We also compare these with the Curzon-Ahlborn efficiency $\eta_{CA}= 1-\sqrt{T_c/T_h}$ and note the curious cross-over between $\eta^*(\mathcal{T})$ and $\eta_{CA}$. The Curzon-Ahlborn efficiency \cite{Curzon-Ahlborn, Novikov} was derived for a specific model of heat transfer---it is not a universal bound. However, as discussed in Ref.~\onlinecite{Review_Stochastic-thermodynamics-fluctuation-theorems-molecular-machines_Seifert_2012}, $\eta_{CA}$ is close to the efficiency at maximum power for many different models. Further discussion on the topic is beyond the scope of this paper and we refer the interested reader to the aforementioned review. Our view is that the problem of maximizing power is very system specific and depends upon the given set-up. To ask for universal bounds on the same requires establishing general features in the model. An attempt along the same direction was made in Ref.~\onlinecite{Eff_Max-Power_Low-Dissipation_Carnot-Engines} where the authors studied a low-dissipation Carnot engine, i.e. one that was operating for a large but finite time period and obtained bounds on the efficiency at maximum power by maximizing power over the thermalization times with the hot and cold reservoirs. (Our problems as studied in \cref{subsubsec:1} and \cref{subsubsec:2} are different since we only optimize over one of the two thermalization times.) They were then able to obtain the Curzon-Ahlborn efficiency as a special case when the dissipation with respect to the reservoirs was symmetric. 
\vspace{-0.5cm}
\section{Summary \& Outlook}\label{sec:Summary}
In this work, we analyzed fluctuations of work done during finite-time processes in two-level systems. We obtained analytic expressions for (a) average work and (b) lower bound for variance as functions of time. We also studied these processes in the context of thermodynamic work extraction cycles performing numerical optimizations for the power output of such cyclic processes. We conclude that finite-time processes are inherently prone to fluctuations that result in broad distributions of work in contrast with what was reported in Ref.~\onlinecite{Baumer2019imperfect}.

Furthermore,  we note that since the Markov process that lies at the heart of the overall physical model is not a simple one, an expression for the variance of work could not be obtained. To illustrate this point, we recall \cref{tab:table6}. It is clear from there that one can write the variance for a discrete $L$-step process as 
\be 
\sigma^2_W = \mathlarger{\sum}_x ~ p(x)W^2(x) - \Big(\mathlarger{\sum}_x ~ p(x) W(x)\Big)^2,
\ee 
where $x$ counts all the paths that correspond to a fixed amount of work $W(x)$ and $p(x)$ is the total probability of occurrence of those paths during the process. For example, there would be various paths corresponding to $W=\epsilon/2$ and one needs to count these paths and sum their contribution which is where the complexity lies. However, there is exactly one path each corresponding to $W=0$ and $W=\epsilon$ respectively. We were able to use this fact to obtain the lower bound for variance. In fact, a similar reasoning was used by the authors of Ref.~\onlinecite{Dhar_PRE2005} to derive the distribution of work for a similar process in the limit of slow driving which allowed them to ignore all but a few relevant paths. Finally, we would like to add that one of the original motivations for this work was to observe the resource resonance phenomenon, as found in Ref.~\onlinecite{Kamil_resonance2019}, within the framework of the thermodynamic resource theory for a physical model. However, we could not make any relevant connections.
\vspace{-0.5cm}
\subsection*{Acknowledgements}

We thank Christopher T. Chubb and Chris Ferrie for helpful discussions during the initial stages of the project. MQ gratefully acknowledges financial support of the Sydney Quantum Academy, Sydney, Australia (proudly funded by the NSW Government) and hospitality of the Theory of Quantum Matter Unit at the Okinawa Institute of Science and Technology, Japan during the writing of this manuscript. KK acknowledges financial support by the Foundation for Polish Science through TEAM-NET project (contract no. POIR.04.04.00-00-17C1/18-00). MT is supported by NUS startup grants (R-263- 000-E32-133 and R-263-000-E32-731) and by the National Research Foundation, Prime Minister’s Office, Singapore and the Ministry of Education, Singapore under the Research Centres of Excellence programme.

\bibliographystyle{apsrev4-1}
\bibliography{draft}

\begin{thebibliography}{35}%
\makeatletter
\providecommand \@ifxundefined [1]{%
 \@ifx{#1\undefined}
}%
\providecommand \@ifnum [1]{%
 \ifnum #1\expandafter \@firstoftwo
 \else \expandafter \@secondoftwo
 \fi
}%
\providecommand \@ifx [1]{%
 \ifx #1\expandafter \@firstoftwo
 \else \expandafter \@secondoftwo
 \fi
}%
\providecommand \natexlab [1]{#1}%
\providecommand \enquote  [1]{``#1''}%
\providecommand \bibnamefont  [1]{#1}%
\providecommand \bibfnamefont [1]{#1}%
\providecommand \citenamefont [1]{#1}%
\providecommand \href@noop [0]{\@secondoftwo}%
\providecommand \href [0]{\begingroup \@sanitize@url \@href}%
\providecommand \@href[1]{\@@startlink{#1}\@@href}%
\providecommand \@@href[1]{\endgroup#1\@@endlink}%
\providecommand \@sanitize@url [0]{\catcode `\\12\catcode `\$12\catcode
  `\&12\catcode `\#12\catcode `\^12\catcode `\_12\catcode `\%12\relax}%
\providecommand \@@startlink[1]{}%
\providecommand \@@endlink[0]{}%
\providecommand \url  [0]{\begingroup\@sanitize@url \@url }%
\providecommand \@url [1]{\endgroup\@href {#1}{\urlprefix }}%
\providecommand \urlprefix  [0]{URL }%
\providecommand \Eprint [0]{\href }%
\providecommand \doibase [0]{http://dx.doi.org/}%
\providecommand \selectlanguage [0]{\@gobble}%
\providecommand \bibinfo  [0]{\@secondoftwo}%
\providecommand \bibfield  [0]{\@secondoftwo}%
\providecommand \translation [1]{[#1]}%
\providecommand \BibitemOpen [0]{}%
\providecommand \bibitemStop [0]{}%
\providecommand \bibitemNoStop [0]{.\EOS\space}%
\providecommand \EOS [0]{\spacefactor3000\relax}%
\providecommand \BibitemShut  [1]{\csname bibitem#1\endcsname}%
\let\auto@bib@innerbib\@empty
\bibitem [{\citenamefont {Callen}(2006)}]{Herbert_thermo}%
  \BibitemOpen
  \bibfield  {author} {\bibinfo {author} {\bibfnamefont {H.~B.}\ \bibnamefont
  {Callen}},\ }\href@noop {} {\emph {\bibinfo {title} {Thermodynamics \& An
  Introduction to Thermostatistics}}}\ (\bibinfo  {publisher} {John Wiley \&
  Sons},\ \bibinfo {year} {2006})\BibitemShut {NoStop}%
\bibitem [{\citenamefont {Jarzynski}(2011)}]{Jarzynski_review}%
  \BibitemOpen
  \bibfield  {author} {\bibinfo {author} {\bibfnamefont {C.}~\bibnamefont
  {Jarzynski}},\ }\href {\doibase 10.1146/annurev-conmatphys-062910-140506}
  {\bibfield  {journal} {\bibinfo  {journal} {Annual Review of Condensed Matter
  Physics}\ }\textbf {\bibinfo {volume} {2}},\ \bibinfo {pages} {329} (\bibinfo
  {year} {2011})}\BibitemShut {NoStop}%
\bibitem [{\citenamefont {Jarzynski}(1997)}]{Jarzynski_PRL1997}%
  \BibitemOpen
  \bibfield  {author} {\bibinfo {author} {\bibfnamefont {C.}~\bibnamefont
  {Jarzynski}},\ }\href {\doibase 10.1103/PhysRevLett.78.2690} {\bibfield
  {journal} {\bibinfo  {journal} {Phys. Rev. Lett.}\ }\textbf {\bibinfo
  {volume} {78}},\ \bibinfo {pages} {2690} (\bibinfo {year}
  {1997})}\BibitemShut {NoStop}%
\bibitem [{\citenamefont {Crooks}(1999)}]{Crooks}%
  \BibitemOpen
  \bibfield  {author} {\bibinfo {author} {\bibfnamefont {G.~E.}\ \bibnamefont
  {Crooks}},\ }\href {\doibase 10.1103/PhysRevE.60.2721} {\bibfield  {journal}
  {\bibinfo  {journal} {Phys. Rev. E}\ }\textbf {\bibinfo {volume} {60}},\
  \bibinfo {pages} {2721} (\bibinfo {year} {1999})}\BibitemShut {NoStop}%
\bibitem [{\citenamefont {Dobson}(2003)}]{Dobson2003}%
  \BibitemOpen
  \bibfield  {author} {\bibinfo {author} {\bibfnamefont {C.~M.}\ \bibnamefont
  {Dobson}},\ }\href {\doibase 10.1038/nature02261} {\bibfield  {journal}
  {\bibinfo  {journal} {Nature}\ }\textbf {\bibinfo {volume} {426}},\ \bibinfo
  {pages} {884} (\bibinfo {year} {2003})}\BibitemShut {NoStop}%
\bibitem [{\citenamefont {{Alemany}}\ and\ \citenamefont
  {{Ritort}}(2010)}]{Alemany&Retort2010}%
  \BibitemOpen
  \bibfield  {author} {\bibinfo {author} {\bibfnamefont {A.}~\bibnamefont
  {{Alemany}}}\ and\ \bibinfo {author} {\bibfnamefont {F.}~\bibnamefont
  {{Ritort}}},\ }\href {\doibase 10.1051/epn/2010205} {\bibfield  {journal}
  {\bibinfo  {journal} {Europhysics News}\ }\textbf {\bibinfo {volume} {41}},\
  \bibinfo {pages} {27} (\bibinfo {year} {2010})}\BibitemShut {NoStop}%
\bibitem [{\citenamefont {Bustamante}\ \emph {et~al.}(2005)\citenamefont
  {Bustamante}, \citenamefont {Liphardt},\ and\ \citenamefont
  {Ritort}}]{Bustamante_2005}%
  \BibitemOpen
  \bibfield  {author} {\bibinfo {author} {\bibfnamefont {C.}~\bibnamefont
  {Bustamante}}, \bibinfo {author} {\bibfnamefont {J.}~\bibnamefont
  {Liphardt}}, \ and\ \bibinfo {author} {\bibfnamefont {F.}~\bibnamefont
  {Ritort}},\ }\href {\doibase 10.1063/1.2012462} {\bibfield  {journal}
  {\bibinfo  {journal} {Physics Today}\ }\textbf {\bibinfo {volume} {58}},\
  \bibinfo {pages} {43–48} (\bibinfo {year} {2005})}\BibitemShut {NoStop}%
\bibitem [{\citenamefont {Kolomeisky}\ and\ \citenamefont
  {Fisher}(2007)}]{fluctuations-biomolecular-review}%
  \BibitemOpen
  \bibfield  {author} {\bibinfo {author} {\bibfnamefont {A.~B.}\ \bibnamefont
  {Kolomeisky}}\ and\ \bibinfo {author} {\bibfnamefont {M.~E.}\ \bibnamefont
  {Fisher}},\ }\href {\doibase 10.1146/annurev.physchem.58.032806.104532}
  {\bibfield  {journal} {\bibinfo  {journal} {Annual Review of Physical
  Chemistry}\ }\textbf {\bibinfo {volume} {58}},\ \bibinfo {pages} {675}
  (\bibinfo {year} {2007})}\BibitemShut {NoStop}%
\bibitem [{\citenamefont {Liphardt}\ \emph {et~al.}(2002)\citenamefont
  {Liphardt}, \citenamefont {Dumont}, \citenamefont {Smith}, \citenamefont
  {Tinoco},\ and\ \citenamefont {Bustamante}}]{Liphardt1832}%
  \BibitemOpen
  \bibfield  {author} {\bibinfo {author} {\bibfnamefont {J.}~\bibnamefont
  {Liphardt}}, \bibinfo {author} {\bibfnamefont {S.}~\bibnamefont {Dumont}},
  \bibinfo {author} {\bibfnamefont {S.~B.}\ \bibnamefont {Smith}}, \bibinfo
  {author} {\bibfnamefont {I.}~\bibnamefont {Tinoco}}, \ and\ \bibinfo {author}
  {\bibfnamefont {C.}~\bibnamefont {Bustamante}},\ }\href {\doibase
  10.1126/science.1071152} {\bibfield  {journal} {\bibinfo  {journal}
  {Science}\ }\textbf {\bibinfo {volume} {296}},\ \bibinfo {pages} {1832}
  (\bibinfo {year} {2002})}\BibitemShut {NoStop}%
\bibitem [{\citenamefont {Garner}(2018)}]{Garner_2018}%
  \BibitemOpen
  \bibfield  {author} {\bibinfo {author} {\bibfnamefont {A.~J.~P.}\
  \bibnamefont {Garner}},\ }\href {\doibase 10.1007/978-3-319-99046-0_27}
  {\bibfield  {journal} {\bibinfo  {journal} {Thermodynamics in the Quantum
  Regime}\ ,\ \bibinfo {pages} {651–679}} (\bibinfo {year}
  {2018})}\BibitemShut {NoStop}%
\bibitem [{\citenamefont {{{\r{A}}berg}}(2013)}]{Aberg-work-extraction_2013}%
  \BibitemOpen
  \bibfield  {author} {\bibinfo {author} {\bibfnamefont {J.}~\bibnamefont
  {{{\r{A}}berg}}},\ }\href {\doibase 10.1038/ncomms2712} {\bibfield  {journal}
  {\bibinfo  {journal} {Nature Communications}\ }\textbf {\bibinfo {volume}
  {4}},\ \bibinfo {eid} {1925} (\bibinfo {year} {2013})}\BibitemShut {NoStop}%
\bibitem [{\citenamefont {Dahlsten}\ \emph {et~al.}(2011)\citenamefont
  {Dahlsten}, \citenamefont {Renner}, \citenamefont {Rieper},\ and\
  \citenamefont {Vedral}}]{Dahlsten_2011}%
  \BibitemOpen
  \bibfield  {author} {\bibinfo {author} {\bibfnamefont {O.~C.~O.}\
  \bibnamefont {Dahlsten}}, \bibinfo {author} {\bibfnamefont {R.}~\bibnamefont
  {Renner}}, \bibinfo {author} {\bibfnamefont {E.}~\bibnamefont {Rieper}}, \
  and\ \bibinfo {author} {\bibfnamefont {V.}~\bibnamefont {Vedral}},\ }\href
  {\doibase 10.1088/1367-2630/13/5/053015} {\bibfield  {journal} {\bibinfo
  {journal} {New Journal of Physics}\ }\textbf {\bibinfo {volume} {13}},\
  \bibinfo {pages} {053015} (\bibinfo {year} {2011})}\BibitemShut {NoStop}%
\bibitem [{\citenamefont {Del~Rio}\ \emph {et~al.}(2011)\citenamefont
  {Del~Rio}, \citenamefont {{\AA}berg}, \citenamefont {Renner}, \citenamefont
  {Dahlsten},\ and\ \citenamefont {Vedral}}]{lidia_2011thermodynamic}%
  \BibitemOpen
  \bibfield  {author} {\bibinfo {author} {\bibfnamefont {L.}~\bibnamefont
  {Del~Rio}}, \bibinfo {author} {\bibfnamefont {J.}~\bibnamefont {{\AA}berg}},
  \bibinfo {author} {\bibfnamefont {R.}~\bibnamefont {Renner}}, \bibinfo
  {author} {\bibfnamefont {O.}~\bibnamefont {Dahlsten}}, \ and\ \bibinfo
  {author} {\bibfnamefont {V.}~\bibnamefont {Vedral}},\ }\href {\doibase
  10.1038/nature10123} {\bibfield  {journal} {\bibinfo  {journal} {Nature}\
  }\textbf {\bibinfo {volume} {474}},\ \bibinfo {pages} {61} (\bibinfo {year}
  {2011})}\BibitemShut {NoStop}%
\bibitem [{\citenamefont {Faist}\ \emph {et~al.}(2015)\citenamefont {Faist},
  \citenamefont {Dupuis}, \citenamefont {Oppenheim},\ and\ \citenamefont
  {Renner}}]{Faist_2015}%
  \BibitemOpen
  \bibfield  {author} {\bibinfo {author} {\bibfnamefont {P.}~\bibnamefont
  {Faist}}, \bibinfo {author} {\bibfnamefont {F.}~\bibnamefont {Dupuis}},
  \bibinfo {author} {\bibfnamefont {J.}~\bibnamefont {Oppenheim}}, \ and\
  \bibinfo {author} {\bibfnamefont {R.}~\bibnamefont {Renner}},\ }\href
  {\doibase 10.1038/ncomms8669} {\bibfield  {journal} {\bibinfo  {journal}
  {Nature Communications}\ }\textbf {\bibinfo {volume} {6}},\  (\bibinfo {year}
  {2015})}\BibitemShut {NoStop}%
\bibitem [{\citenamefont {Tomamichel}(2015)}]{tomamichel2015quantum}%
  \BibitemOpen
  \bibfield  {author} {\bibinfo {author} {\bibfnamefont {M.}~\bibnamefont
  {Tomamichel}},\ }\href {\doibase 10.1007/978-3-319-21891-5} {\emph {\bibinfo
  {title} {Quantum Information Processing with Finite Resources: Mathematical
  Foundations}}},\ Vol.~\bibinfo {volume} {5}\ (\bibinfo  {publisher}
  {Springer},\ \bibinfo {year} {2015})\BibitemShut {NoStop}%
\bibitem [{\citenamefont {Rényi}(1961)}]{renyi1961}%
  \BibitemOpen
  \bibfield  {author} {\bibinfo {author} {\bibfnamefont {A.}~\bibnamefont
  {Rényi}},\ }in\ \href {https://projecteuclid.org/euclid.bsmsp/1200512181}
  {\emph {\bibinfo {booktitle} {Proceedings of the Fourth Berkeley Symposium on
  Mathematical Statistics and Probability, Volume 1: Contributions to the
  Theory of Statistics}}}\ (\bibinfo  {publisher} {University of California
  Press},\ \bibinfo {year} {1961})\BibitemShut {NoStop}%
\bibitem [{\citenamefont {{Renner}}\ and\ \citenamefont
  {{Wolf}}(2004)}]{renner-wolf2004}%
  \BibitemOpen
  \bibfield  {author} {\bibinfo {author} {\bibfnamefont {R.}~\bibnamefont
  {{Renner}}}\ and\ \bibinfo {author} {\bibfnamefont {S.}~\bibnamefont
  {{Wolf}}},\ }in\ \href {\doibase 10.1109/ISIT.2004.1365269} {\emph {\bibinfo
  {booktitle} {International Symposium on Information Theory, 2004. ISIT 2004.
  Proceedings.}}}\ (\bibinfo {year} {2004})\BibitemShut {NoStop}%
\bibitem [{\citenamefont {{Renner}}(2005)}]{renner_phd-thesis2005}%
  \BibitemOpen
  \bibfield  {author} {\bibinfo {author} {\bibfnamefont {R.}~\bibnamefont
  {{Renner}}},\ }\emph {\bibinfo {title} {Security of Quantum Key
  Distribution}},\ \href@noop {} {Ph.D. thesis},\ \bibinfo  {school} {ETH
  Zurich} (\bibinfo {year} {2005}),\ \Eprint
  {http://arxiv.org/abs/arXiv:quant-ph/0512258v2} {arXiv:quant-ph/0512258v2}
  \BibitemShut {NoStop}%
\bibitem [{\citenamefont {Ziman}\ \emph
  {et~al.}(2005{\natexlab{a}})\citenamefont {Ziman}, \citenamefont {{\v
  S}telmachovi{\v c}},\ and\ \citenamefont {Bu{\v
  z}ek}}]{CollisionModel_OpenQuantumDynamics2005}%
  \BibitemOpen
  \bibfield  {author} {\bibinfo {author} {\bibfnamefont {M.}~\bibnamefont
  {Ziman}}, \bibinfo {author} {\bibfnamefont {P.}~\bibnamefont {{\v
  S}telmachovi{\v c}}}, \ and\ \bibinfo {author} {\bibfnamefont
  {V.}~\bibnamefont {Bu{\v z}ek}},\ }\href {\doibase 10.1007/s11080-005-0488-0}
  {\bibfield  {journal} {\bibinfo  {journal} {Open Systems \& Information
  Dynamics}\ }\textbf {\bibinfo {volume} {12}},\ \bibinfo {pages} {81}
  (\bibinfo {year} {2005}{\natexlab{a}})}\BibitemShut {NoStop}%
\bibitem [{\citenamefont {Scarani}\ \emph {et~al.}(2002)\citenamefont
  {Scarani}, \citenamefont {Ziman}, \citenamefont {\ifmmode \check{S}\else
  \v{S}\fi{}telmachovi\ifmmode~\check{c}\else \v{c}\fi{}}, \citenamefont
  {Gisin},\ and\ \citenamefont {Bu\ifmmode~\check{z}\else
  \v{z}\fi{}ek}}]{CollisionModel_ThermalizingQuantumMachines_PRL2002}%
  \BibitemOpen
  \bibfield  {author} {\bibinfo {author} {\bibfnamefont {V.}~\bibnamefont
  {Scarani}}, \bibinfo {author} {\bibfnamefont {M.}~\bibnamefont {Ziman}},
  \bibinfo {author} {\bibfnamefont {P.}~\bibnamefont {\ifmmode \check{S}\else
  \v{S}\fi{}telmachovi\ifmmode~\check{c}\else \v{c}\fi{}}}, \bibinfo {author}
  {\bibfnamefont {N.}~\bibnamefont {Gisin}}, \ and\ \bibinfo {author}
  {\bibfnamefont {V.}~\bibnamefont {Bu\ifmmode~\check{z}\else \v{z}\fi{}ek}},\
  }\href {\doibase 10.1103/PhysRevLett.88.097905} {\bibfield  {journal}
  {\bibinfo  {journal} {Phys. Rev. Lett.}\ }\textbf {\bibinfo {volume} {88}},\
  \bibinfo {pages} {097905} (\bibinfo {year} {2002})}\BibitemShut {NoStop}%
\bibitem [{\citenamefont {Ciccarello}(2017)}]{Collisionmodels_quantumoptics}%
  \BibitemOpen
  \bibfield  {author} {\bibinfo {author} {\bibfnamefont {F.}~\bibnamefont
  {Ciccarello}},\ }\href {\doibase https://doi.org/10.1515/qmetro-2017-0007}
  {\bibfield  {journal} {\bibinfo  {journal} {Quantum Measurements and Quantum
  Metrology}\ }\textbf {\bibinfo {volume} {4}},\ \bibinfo {pages} {53 }
  (\bibinfo {year} {2017})}\BibitemShut {NoStop}%
\bibitem [{\citenamefont {Marathe}\ and\ \citenamefont
  {Dhar}(2005)}]{Dhar_PRE2005}%
  \BibitemOpen
  \bibfield  {author} {\bibinfo {author} {\bibfnamefont {R.}~\bibnamefont
  {Marathe}}\ and\ \bibinfo {author} {\bibfnamefont {A.}~\bibnamefont {Dhar}},\
  }\href {\doibase 10.1103/PhysRevE.72.066112} {\bibfield  {journal} {\bibinfo
  {journal} {Phys. Rev. E}\ }\textbf {\bibinfo {volume} {72}},\ \bibinfo
  {pages} {066112} (\bibinfo {year} {2005})}\BibitemShut {NoStop}%
\bibitem [{\citenamefont {B{\"{a}}umer}\ \emph {et~al.}(2019)\citenamefont
  {B{\"{a}}umer}, \citenamefont {Perarnau-Llobet}, \citenamefont
  {Kammerlander}, \citenamefont {Wilming},\ and\ \citenamefont
  {Renner}}]{Baumer2019imperfect}%
  \BibitemOpen
  \bibfield  {author} {\bibinfo {author} {\bibfnamefont {E.}~\bibnamefont
  {B{\"{a}}umer}}, \bibinfo {author} {\bibfnamefont {M.}~\bibnamefont
  {Perarnau-Llobet}}, \bibinfo {author} {\bibfnamefont {P.}~\bibnamefont
  {Kammerlander}}, \bibinfo {author} {\bibfnamefont {H.}~\bibnamefont
  {Wilming}}, \ and\ \bibinfo {author} {\bibfnamefont {R.}~\bibnamefont
  {Renner}},\ }\href {\doibase 10.22331/q-2019-06-24-153} {\bibfield  {journal}
  {\bibinfo  {journal} {{Quantum}}\ }\textbf {\bibinfo {volume} {3}},\ \bibinfo
  {pages} {153} (\bibinfo {year} {2019})}\BibitemShut {NoStop}%
\bibitem [{\citenamefont {Brun}(2002)}]{CollisionModel_LME_ToddBrun}%
  \BibitemOpen
  \bibfield  {author} {\bibinfo {author} {\bibfnamefont {T.~A.}\ \bibnamefont
  {Brun}},\ }\href {\doibase 10.1119/1.1475328} {\bibfield  {journal} {\bibinfo
   {journal} {American Journal of Physics}\ }\textbf {\bibinfo {volume} {70}},\
  \bibinfo {pages} {719} (\bibinfo {year} {2002})}\BibitemShut {NoStop}%
\bibitem [{\citenamefont {Ziman}\ \emph
  {et~al.}(2005{\natexlab{b}})\citenamefont {Ziman}, \citenamefont
  {Štelmachovič},\ and\ \citenamefont
  {Bužek}}]{OpenSystem_CollisionModel_Ziman}%
  \BibitemOpen
  \bibfield  {author} {\bibinfo {author} {\bibfnamefont {M.}~\bibnamefont
  {Ziman}}, \bibinfo {author} {\bibfnamefont {P.}~\bibnamefont
  {Štelmachovič}}, \ and\ \bibinfo {author} {\bibfnamefont {V.}~\bibnamefont
  {Bužek}},\ }\href {\doibase 10.1007/s11080-005-0488-0} {\bibfield  {journal}
  {\bibinfo  {journal} {Open Systems \& Information Dynamics}\ }\textbf
  {\bibinfo {volume} {12}},\ \bibinfo {pages} {81} (\bibinfo {year}
  {2005}{\natexlab{b}})}\BibitemShut {NoStop}%
\bibitem [{\citenamefont {Tange}(2018)}]{gnu_parallel}%
  \BibitemOpen
  \bibfield  {author} {\bibinfo {author} {\bibfnamefont {O.}~\bibnamefont
  {Tange}},\ }\href {https://dl.acm.org/doi/book/10.5555/3235180} {\emph
  {\bibinfo {title} {GNU Parallel 2018}}}\ (\bibinfo  {publisher} {Lulu.com},\
  \bibinfo {year} {2018})\BibitemShut {NoStop}%
\bibitem [{\citenamefont {{Quan}}\ \emph {et~al.}(2007)\citenamefont {{Quan}},
  \citenamefont {{Liu}}, \citenamefont {{Sun}},\ and\ \citenamefont
  {{Nori}}}]{Quan_qthermo-cycles-I2007}%
  \BibitemOpen
  \bibfield  {author} {\bibinfo {author} {\bibfnamefont {H.~T.}\ \bibnamefont
  {{Quan}}}, \bibinfo {author} {\bibfnamefont {Y.-X.}\ \bibnamefont {{Liu}}},
  \bibinfo {author} {\bibfnamefont {C.~P.}\ \bibnamefont {{Sun}}}, \ and\
  \bibinfo {author} {\bibfnamefont {F.}~\bibnamefont {{Nori}}},\ }\href
  {\doibase 10.1103/PhysRevE.76.031105} {\bibfield  {journal} {\bibinfo
  {journal} {\pre}\ }\textbf {\bibinfo {volume} {76}},\ \bibinfo {eid} {031105}
  (\bibinfo {year} {2007})}\BibitemShut {NoStop}%
\bibitem [{Note1()}]{Note1}%
  \BibitemOpen
  \bibinfo {note} {This means that if the system was in the ground state then
  it continues to be in the ground state of the new Hamiltonian.}\BibitemShut
  {Stop}%
\bibitem [{\citenamefont
  {Seifert}(2012)}]{Review_Stochastic-thermodynamics-fluctuation-theorems-molecular-machines_Seifert_2012}%
  \BibitemOpen
  \bibfield  {author} {\bibinfo {author} {\bibfnamefont {U.}~\bibnamefont
  {Seifert}},\ }\href {\doibase 10.1088/0034-4885/75/12/126001} {\bibfield
  {journal} {\bibinfo  {journal} {Reports on Progress in Physics}\ }\textbf
  {\bibinfo {volume} {75}},\ \bibinfo {pages} {126001} (\bibinfo {year}
  {2012})}\BibitemShut {NoStop}%
\bibitem [{\citenamefont {{Wolfram Research, Inc.}}()}]{Mathematica}%
  \BibitemOpen
  \bibfield  {author} {\bibinfo {author} {\bibnamefont {{Wolfram Research,
  Inc.}}},\ }\href {https://www.wolfram.com/mathematica/} {\enquote {\bibinfo
  {title} {Mathematica, {V}ersion 12.1},}\ }\bibinfo {note} {Champaign, IL,
  2020}\BibitemShut {NoStop}%
\bibitem [{\citenamefont {{Wolfram Language \& System Documentation
  Center}}()}]{Mathematica-ref}%
  \BibitemOpen
  \bibfield  {author} {\bibinfo {author} {\bibnamefont {{Wolfram Language \&
  System Documentation Center}}},\ }\href
  {https://reference.wolfram.com/language/tutorial/ConstrainedOptimizationGlobalNumerical.html}
  {\enquote {\bibinfo {title} {Constrained optimization},}\ }\BibitemShut
  {NoStop}%
\bibitem [{\citenamefont {{Curzon}}\ and\ \citenamefont
  {{Ahlborn}}(1975)}]{Curzon-Ahlborn}%
  \BibitemOpen
  \bibfield  {author} {\bibinfo {author} {\bibfnamefont {F.~L.}\ \bibnamefont
  {{Curzon}}}\ and\ \bibinfo {author} {\bibfnamefont {B.}~\bibnamefont
  {{Ahlborn}}},\ }\href {\doibase 10.1119/1.10023} {\bibfield  {journal}
  {\bibinfo  {journal} {American Journal of Physics}\ }\textbf {\bibinfo
  {volume} {43}},\ \bibinfo {pages} {22} (\bibinfo {year} {1975})}\BibitemShut
  {NoStop}%
\bibitem [{\citenamefont {Novikov}(1958)}]{Novikov}%
  \BibitemOpen
  \bibfield  {author} {\bibinfo {author} {\bibfnamefont {I.}~\bibnamefont
  {Novikov}},\ }\href {\doibase https://doi.org/10.1016/0891-3919(58)90244-4}
  {\bibfield  {journal} {\bibinfo  {journal} {Journal of Nuclear Energy
  (1954)}\ }\textbf {\bibinfo {volume} {7}},\ \bibinfo {pages} {125 } (\bibinfo
  {year} {1958})}\BibitemShut {NoStop}%
\bibitem [{\citenamefont {Esposito}\ \emph {et~al.}(2010)\citenamefont
  {Esposito}, \citenamefont {Kawai}, \citenamefont {Lindenberg},\ and\
  \citenamefont {Van~den
  Broeck}}]{Eff_Max-Power_Low-Dissipation_Carnot-Engines}%
  \BibitemOpen
  \bibfield  {author} {\bibinfo {author} {\bibfnamefont {M.}~\bibnamefont
  {Esposito}}, \bibinfo {author} {\bibfnamefont {R.}~\bibnamefont {Kawai}},
  \bibinfo {author} {\bibfnamefont {K.}~\bibnamefont {Lindenberg}}, \ and\
  \bibinfo {author} {\bibfnamefont {C.}~\bibnamefont {Van~den Broeck}},\ }\href
  {\doibase 10.1103/PhysRevLett.105.150603} {\bibfield  {journal} {\bibinfo
  {journal} {Phys. Rev. Lett.}\ }\textbf {\bibinfo {volume} {105}},\ \bibinfo
  {pages} {150603} (\bibinfo {year} {2010})}\BibitemShut {NoStop}%
\bibitem [{\citenamefont {Korzekwa}\ \emph {et~al.}(2019)\citenamefont
  {Korzekwa}, \citenamefont {Chubb},\ and\ \citenamefont
  {Tomamichel}}]{Kamil_resonance2019}%
  \BibitemOpen
  \bibfield  {author} {\bibinfo {author} {\bibfnamefont {K.}~\bibnamefont
  {Korzekwa}}, \bibinfo {author} {\bibfnamefont {C.~T.}\ \bibnamefont {Chubb}},
  \ and\ \bibinfo {author} {\bibfnamefont {M.}~\bibnamefont {Tomamichel}},\
  }\href {\doibase 10.1103/PhysRevLett.122.110403} {\bibfield  {journal}
  {\bibinfo  {journal} {Phys. Rev. Lett.}\ }\textbf {\bibinfo {volume} {122}},\
  \bibinfo {pages} {110403} (\bibinfo {year} {2019})}\BibitemShut {NoStop}%
\end{thebibliography}%


%
\onecolumngrid
\begin{appendix}
\section{Quantum Isothermal processes}\label{app:2}
In the following lemma, we denote the change in the energy of a system during a thermodynamic process $a\mapsto b$ by $\Delta U_{ab}$, the heat exchanged by $Q_{ab}$ and the work done by $W_{ab}$.

\begin{lemma}
A quantum isothermal expansion is such that the gaps between the energy levels of the Hamiltonian $H$ are scaled by a factor $k<1$.
\end{lemma}
\bpr
A-priori, there is nothing constraining the value of $k$ other than the trivial requirement of $k>0$. However, it is clear that if we choose $k>1$ then we are stretching the energy levels apart while if $k<1$ then we are compressing them together. For an isothermal process ($a\mapsto b$) to be an expansion, the following should be true
\begin{flalign*}
Q_{ab}>0\iff \Delta U_{ab}&> W_{ab}.
\end{flalign*}
We know that $ W_{ab}=\Delta F_{ab}=T_h\ln{\big(Z_a/Z_b\big)}$ \cite{Aberg-work-extraction_2013}, where $Z_a=\sum_ie^{-\epsilon_i/T_h}$ and $Z_b=\sum_ie^{-k\epsilon_i/T_h}$. Therefore, the following inequality must be satisfied by any quantum isothermal expansion:
\be\label{iso-exp}
\sum_j \epsilon_j\Bigg(k\frac{e^{-k\epsilon_j/T_h}}{Z_b}-\frac{e^{-\epsilon_j/T_h}}{Z_a}\Bigg)>T_h\Big(\ln{Z_a}-\ln{Z_b}\Big).
\ee
Let us assume that $k>1$, then 
\be
e^{-k\epsilon_i/T_h}&<&e^{-\epsilon_i/T_h}\label{eq14}\\
\implies Z_b&<&Z_a.\label{eq15}
\ee
Thus, \eqref{iso-exp} implies
\benn
\sum_j \epsilon_j\Bigg(k\frac{e^{-k\epsilon_j/T_h}}{Z_b}-\frac{e^{-\epsilon_j/T_h}}{Z_a}\Bigg)&>&0,\eenn
which implies
\be\label{eq16}
k\frac{e^{-k\epsilon_j/T_h}}{Z_b}>\frac{e^{-\epsilon_j/T_h}}{Z_a}.
\ee
Now, \eqref{eq15} implies
\be\label{eq17a}
\frac{e^{-k\epsilon_i/T_h}}{Z_b}&>&\frac{e^{-k\epsilon_i/T_h}}{Z_a},
\ee
but, by \eqref{eq14}, one has
\be\label{eq17b}
\frac{e^{-\epsilon_i/T_h}}{Z_b}>\frac{e^{-k\epsilon_i/T_h}}{Z_b}.
\ee
Combining \eqref{eq17a} with \eqref{eq17b}, we have
\be
\frac{e^{-\epsilon_i/T_h}}{Z_b}>\frac{e^{-k\epsilon_i/T_h}}{Z_a}.
\ee 
Differentiating both sides with respect to $\epsilon_i$, we obtain
\benn
\frac{e^{-\epsilon_i/T_h}}{Z_b}&<&\frac{ke^{-k\epsilon_i/T_h}}{Z_a}.
\eenn
But, this contradicts \eqref{eq16}. So, the assumption is wrong which implies that $k<1$ for a quantum isothermal expansion.
\epr

\section{Aside on special functions}\label{app:special_functions}
\begin{definition*}[Hypergeometric function in integral form]
\be\label{def:eq:general integral_hypergeometric}
&~&\bigintsss \D x \frac{e^{px}}{1+e^{(q-rx)}}=\notag \\
&~&\frac{e^{(p+r)x+q}}{p+r}\times
~_2F_1\Bigg(1, \frac{p}{r}+1, \frac{p}{r}+2; -e^{-(q-rx)}\Bigg),
\ee
where $p,q$, and $r$ are rationals and the hypergeometric function is
\be\label{def:eq:hypergeometric_as_series}
_2F_1: (a,b,c, z)\mapsto \mathlarger{\mathlarger{\sum}}_{n=0}^\infty\frac{(a)_n (b)_n}{(c)_n} \frac{z^n}{n!},~~~~|z|\leq1,
\ee
and $(x)_n$ denotes the \emph{rising factorial}
\be
(x)_n=\begin{cases}
                    1,~~~~n=0,\\
                    x(x+1)\cdots(x+n-1),~~~~~n>0.
     \end{cases}                
\ee
\end{definition*}
\begin{definition*}[Lerch transcendent]\label{def:Lerchi_Phi}
\be
\Phi_L : (z,s,a) \mapsto \mathlarger{\mathlarger{\sum}}_{n=0}^\infty\frac{z^n}{(n+a)^s}.
\ee 
where $z\in\mathbb{C}$ and $Re(a)>0$. We can then write the function $\mathcal{G}$ (\cref{def:G}) in terms of the Lerch transcendent as 
\be\label{eq:def:G2}
	\mathcal{G}(t) = \frac{\kappa\tau T_h}{\epsilon}e^{-\delta(t)/T_h}~\Phi_L(-e^{-\delta(t)/T_h},1,\frac{\kappa\tau T_h}{\epsilon}+1).
\ee 
\end{definition*}

\section{Proof of \cref{lemma:Carnot_cycle_optimal}}\label{app:3}
\bpr
Note that the points $a,b,c,$ and $d$ as in \cref{fig:CarnotCycle}, defining a cycle of a Carnot engine,  are not independent---see \eqref{eq:carnot_cycle:1} and \eqref{eq:carnot_cycle:2}. Thus, there are only two free variables that define any particular cycle. Let us set $\delta_a$ and $\delta_c$ as the independent ones. Then, changing variables in \eqref{eq:avg_work_C} and plugging the expressions for the partition function $Z$, $p_a$ and $p_b$, we obtain 
\be\label{eq:lemma:carnot_cycle:1}
	\mu_{W_C}(\delta_a,~\delta_c) &=& \Big(T_h - T_c\Big)\Bigg(\ln\bigg(\frac{1+e^{-\delta_a/T_h}}{1+e^{-\delta_c/T_c}}\bigg) + \frac{\delta_a/T_h}{1+e^{\delta_a/T_h}} - \frac{\delta_c/T_c}{1+e^{\delta_c/T_c}}\Bigg).
\ee 
Introducing $x:=\delta_a/T_h$ and $y:=\delta_c/T_c$ reduces the equation above to  
\be\label{eq:lemma:carnot_cycle:2}
	\mu_{W_C}(x,~y) = \Big(T_h - T_c\Big)\Bigg(\ln\bigg(\frac{1+e^{-x}}{1+e^{-y}}\bigg) +  \frac{x}{1+e^{x}} - \frac{y}{1+e^{y}}\Bigg).
	\ee 
	
Let us now look at the function 
\be 
f:x\mapsto  (1+e^{-x})~e^{\frac{x}{1+e^{x}}}.
\ee 	
Evaluating the derivative of this function we obtain 
\be 
f'(x) = - \bigg(\frac{x}{1+e^x}\bigg)~e^{\frac{x}{1+e^{x}}} <0,~~~~\forall~x>0.
\ee 
This means that $f$ is a monotonically decreasing function on $\mathbb{R}^+$. Hence, the minimum/maximum would be attained on the boundaries of the interval $\mathcal{I} \subset \mathbb{R}^+$. Note that \eqref{eq:lemma:carnot_cycle:2} can be written in terms of $f$ simply as 
\be 
\mu_{W_C}(x,~y) = \Big(T_h - T_c\Big)\Big( \ln~f(x) - \ln~f(y)\Big).
\ee 
As $\ln$ is a monotonically increasing function, $\ln\circ f$ would thus be monotonically decreasing since $f$ is monotonically decreasing. Now, as $\mu_{W_C}<0$ where the negative sign implies work output, maximizing the average work output amounts to minimizing $\mu_{W_C}$ with respect to $x$ and $y$. So, we have
\be 
\min_{x,~y\in\mathcal{I}} \mu_{W_C} (x,~y) &=& \min_{x,~y\in\mathcal{I}} \Big(T_h - T_c\Big)\Big( \ln~f(x) - \ln~f(y)\Big)\\
&=& \Big(T_h - T_c\Big) \Big(  \min_{x\in\mathcal{I}}\ln~f(x) -  \max_{y\in\mathcal{I}}\ln~f(y)\Big).
\ee 
As $\ln\circ f$ is monotonically decreasing it implies that 
\be 
\min_{x,~y\in\mathcal{I}} \mu_{W_C} (x,~y) &=& \Big(T_h - T_c\Big) \Bigg( \ln~f\Big(\max_{x\in\mathcal{I}} x\Big) - \ln~f\Big(\min_{y\in\mathcal{I}} y\Big)\Bigg).
\ee 
Substituting for $x$ and $y$ in terms in of $\delta_a$ and $\delta_c$ and noting that
\be
\max_{\delta_a}\delta_a = \delta_{\max},~~~~\textrm{and}~~\min_{\delta_c}\delta_c = \delta_{\min},
\ee 
gives us
\be 
\argmax_{\delta_a,~\delta_c}\mu_{W_C}(\delta_a,\delta_c) = (\delta_{\max},~\delta_{\min}).
\ee 
\epr
\section{Partial thermalization under \cref{assumption:1} while increasing the energy gap}\label{app:4}
\begin{definition*} Given the energy gap $\delta(t)$ of a two-level system at time $t<\tau$, we define the function
	\be \label{def:H}
		\mathcal{H}: t \mapsto \mathlarger{\mathlarger{\mathlarger{\sum}}}_{n=0}^\infty \frac{\Big(e^{-\frac{\delta(t)}{T_c}}\Big)^n}{\Big(\frac{n\epsilon}{\kappa\tau T_c}+1\Big)},
	\ee 
	where $\epsilon=\delta_{\max}-\delta_{\min}$, $\kappa$ is the thermalization rate, and $T_c$ is the temperature of the ambient bath.
\end{definition*}

The function $\mathcal{H}$ is a monotone function in $t$. For $\delta$ monotonically increasing in $t$, $\mathcal{H}$ monotonically decreases. This follows by noting that $e^{-\delta}$ is also monotonically decreasing in $t$.

\begin{lemma}[Time evolution of occupation probabilities under partial thermalization while increasing the energy gap]\label{lemma:p(t)_rev}
Given a two-level system that undergoes partial thermalization as per \cref{assumption:1} in the presence of a bath at temperature $T_c$ for a time $\tau$ such that its energy gap changes from $\delta_{min}$ to $\delta_{max}$, the probability of the system to be in the excited state at any time $0<t<\tau$ is 
\be \label{eq:lemma:p(t)_rev}
p(t) &=& p_0 e^{-\kappa t} +\mathcal{H}(t) - e^{-\kappa t}\mathcal{H}(0),
\ee 	
where $\epsilon =\delta_{max}-\delta_{min}$, $p_0 = p(0)$ and $\delta(t) = \delta_{min} + \epsilon t /\tau$.
\end{lemma}
\bpr
Re-writing the differential equation for general partial thermalization processes where the hot bath is replaced by the cold bath in \eqref{eq:lemma:PTh_dp/dt}, we have
\be
\frac{\D p}{\D t} + \kappa p(t) = \kappa \gamma_c\big(\delta(t)\big),
\ee
which can be integrated along with the initial condition $p(0)=p_0$ to obtain 
\be\label{eq:lemma:p(t)_rev:1}
p(t) = p_0 e^{-\kappa t} + \kappa e^{-\kappa t}\int_0^t e^{\kappa t'} \gamma_c\big(\delta(t')\big) ~\D t'.
\ee
Given \cref{assumption:1} and the boundary conditions $\delta(0)=\delta_{\min}$ and $\delta(\tau)=\delta_{\max}$, we have 
\be\label{eq:lemma:p(t)_rev:3}
\delta(t)=\delta_{\min} + \frac{\epsilon}{\tau} t,
\ee
where $\epsilon = \delta_{\max}-\delta_{\min}$. Plugging
$
\gamma_c\big(\delta(t)\big)=\frac{1}{1+e^{\delta(t)/T_c}}
$ and \eqref{eq:lemma:p(t)_rev:3} in \eqref{eq:lemma:p(t)_rev:1}, we obtain 
\be \label{eq:lemma:p(t)_rev:4}
p(t) = p_0 e^{-\kappa t} + \kappa e^{-\kappa t}\bigints_0^t \frac{e^{\kappa t'}}{1+e^{\frac{(\delta_{\min } + \epsilon t'/\tau)}{T_c}}} ~\D t'.
\ee 
Evaluating the integral above, we obtain 
\be\label{eq:lemma:p(t)_rev:5}
p(t) &=& p_0 e^{-\kappa t} + \kappa e^{-\kappa t}\Bigg\{\frac{e^{\kappa t'}}{\kappa}~_2F_1\Big(1, \frac{\kappa \tau T_c}{\epsilon}, \frac{\kappa \tau T_c}{\epsilon}+1; -e^{\frac{(\delta_{\min} + \epsilon t'/\tau)}{T_c}}\Big)\Bigg|_0^t \notag\Bigg\}\\
&=& p_0 e^{-\kappa t} + \kappa e^{-\kappa t}\Bigg\{\frac{e^{\kappa t}}{\kappa}~_2F_1\Big(1, \frac{\kappa \tau T_c}{\epsilon}, \frac{\kappa \tau T_c}{\epsilon}+1; -e^{\frac{\delta(t)}{T_c}}\Big) - \frac{1}{\kappa}~_2F_1\Big(1, \frac{\kappa \tau T_c}{\epsilon}, \frac{\kappa \tau T_c}{\epsilon}+1; -e^{\frac{\delta_{\min}}{T_c}}\Big)\Bigg\} \notag \\
&=& p_0 e^{-\kappa t} + ~_2F_1\Big(1, \frac{\kappa \tau T_c}{\epsilon}, \frac{\kappa \tau T_c}{\epsilon}+1; -e^{\frac{\delta(t)}{T_c}}\Big) - e^{-\kappa t}~_2F_1\Big(1, \frac{\kappa \tau T_c}{\epsilon}, \frac{\kappa \tau T_c}{\epsilon}+1; -e^{\frac{\delta_{\min}}{T_c}}\Big).
\ee

Next, using \cref{def:general integral_hypergeometric} we can write 
\be\label{eq:lemma:p(t)_rev:6}
 ~_2F_1(1, a, 1+a; -z) &=& \mathlarger{\mathlarger{\sum}}_{n=0}^\infty \frac{n! (a)_n}{(1+a)_n} \frac{(-z)^n}{n!}\notag\\
&=& \mathlarger{\mathlarger{\sum}}_{n=0}^\infty \frac{(a)(1+a)\cdots (n-1 +a)}{(1+a)(2+a)\cdots (n+a)} (-z)^n \notag\\
&=& \mathlarger{\mathlarger{\sum}}_{n=0}^\infty \frac{~a~(-z)^{n}}{(n+a)} \notag \\
&=& \mathlarger{\mathlarger{\sum}}_{n=0}^\infty \frac{(-z)^{n}}{(\frac{n}{a}+1)}.
\ee
Using \eqref{eq:lemma:p(t)_rev:6} we can write \eqref{eq:lemma:p(t)_rev:5} in terms of $\mathcal{H}$ to obtain \eqref{eq:lemma:p(t)_rev}.
\epr

\begin{lemma}[Average work when increasing the energy gap]\label{lemma:average work_rev}

 The average work done by a two-level system during a process as per \cref{assumption:1} along with partial thermalizations in the presence of a bath at temperature $T_h$ for a time $\tau$ such that its energy gap changes from $\delta_{\min}$ to $\delta_{\max}$ is
 
\be\label{eq:lemma:<W>_rev}
\mu_W(\tau)  &=& - W_{iso}^{T_c} -  \frac{W_{ad}}{\kappa \tau}\big(1-e^{-\kappa \tau}\big) + \frac{\epsilon}{\kappa \tau}\Bigg\{\mathcal{H}(\tau) -  e^{-\kappa \tau}\mathcal{H}(0)\Bigg\},
\ee
where $\epsilon =\delta_{max}-\delta_{min}$, $p_0 = p(0)$, and $W_{iso}^{T_c}$ is the work output of the corresponding isothermal process, i.e. $W_{iso}^{T_c}=-T_c\log{\frac{Z(\delta_{\min})}{Z(\delta_{\max})}}$, where $Z$ is the partition function $Z: \delta \mapsto 1+e^{-\delta/T_c}$.
\end{lemma}

\bpr
We start by noting that 
\be\label{eq:lemma:<W>_rev:1}
\frac{\D p}{\D \delta}&=&\frac{\D p}{\D t}.\frac{\D t}{\D \delta}\notag\\
&=& -\frac{\kappa \tau}{\epsilon}\Big(\gamma_h(\delta)-p\Big), 
\ee
where the last line follows from \eqref{eq:lemma:PTh_dp/dt} and \eqref{eq:PTh_ddelta/dt} while supressing the dependence on $t$. Integrating \eqref{eq:lemma:<W>:1} with respect to $\delta$ from $\delta_{\min }$ to $\delta_{\max }$, we have
\be\label{eq:lemma:<W>_rev:2a}
\bigintssss_{\delta_{\min }}^{\delta_{\max }} p~\D \delta = \bigintssss_{\delta_{\min }}^{\delta_{\max }}\gamma_c\big(\delta\big)\D \delta +  \frac{\epsilon}{\kappa \tau}\bigintssss_{\delta_{\min }}^{\delta_{\max }} \frac{\D p}{\D \delta}\D \delta.
\ee
Thus, \eqref{eq:PTh_<W>_cts} and \eqref{eq:lemma:<W>_rev:2a} together imply
\be\label{eq:lemma:<W>_rev:2b}
\mu_W(\tau)= \bigintssss_{\delta_{\min }}^{\delta_{\max }}\gamma_c\big(\delta\big)\D \delta + \frac{\epsilon}{\kappa \tau}\bigintssss_{\delta_{\min }}^{\delta_{\max }} \frac{\D p}{\D \delta}\D \delta.
\ee 
The first term on the right-hand side is the negative of the work done during the corresponding isothermal process (when energy gap changes from $\delta_{\max}$ to $\delta_{\min}$). Substituting the expression for $\gamma_c(\delta)$ and evaluating the integral gives us the first term of \eqref{eq:lemma:<W>_rev:2b} as
\be\label{eq:lemma:<W>_rev:3}
\bigintssss_{\delta_{\min}}^{\delta_{\max}}\gamma_c\big(\delta\big) \D \delta = T_c \log\frac{Z(\delta_{\min})}{Z(\delta_{\max})}\triangleq - W_{iso}^{T_c},
\ee
where $Z$ is the partition function $Z : t \mapsto 1+e^{-\delta(t)/T_c}$.
Next, we evaluate the integral in the second term in \eqref{eq:lemma:<W>_rev:2b} using \cref{lemma:p(t)_rev}. First we note that
\benn 
\bigintssss_{\delta_{\min}}^{\delta_{\max }} \frac{\D p}{\D \delta}~\D \delta &=& p(\delta_{\max})-p(\delta_{\min }).
\eenn
As $p(\delta_{\min })=p(0)=p_0$ is given and $p(\delta_{\max })=p(\tau)$, we use \eqref{eq:lemma:p(t)_rev} to obtain
\be\label{eq:lemma:<W>_rev:4}
p(\delta_{\max })-p(\delta_{\min }) &=& -p_0(1-e^{-\kappa \tau}) +\mathcal{H}(\tau) - e^{-\kappa \tau}\mathcal{H}(0).
\ee
Plugging \eqref{eq:lemma:<W>_rev:3} and \eqref{eq:lemma:<W>_rev:4} in \eqref{eq:lemma:<W>_rev:2b} along with \eqref{eq:lemma:<W>:5} we obtain \eqref{eq:lemma:<W>_rev}.
\epr 

\end{appendix}
\end{document}